\documentclass[sigconf, nonacm=true]{acmart}

\usepackage{multirow}
\usepackage{algpseudocode}
\usepackage{algorithm}
\usepackage{graphicx}
\usepackage{subcaption}
\usepackage{array}
\usepackage[export]{adjustbox}
\usepackage{flushend}

\usepackage{listings}               % for code listings
\usepackage{xcolor}                 % for styling code

\definecolor{backcolour}{RGB}{246, 246, 246}   % 0xF6, 0xF6, 0xF6
\definecolor{codegreen}{RGB}{16, 124, 2}       % 0x10, 0x7C, 0x02
\definecolor{codepurple}{RGB}{170, 0, 217}     % 0xAA, 0x00, 0xD9
\definecolor{codered}{RGB}{154, 0, 18}         % 0x9A, 0x00, 0x12

\lstdefinestyle{gcolabstyle}{
  basicstyle=\linespread{1}\ttfamily\small\selectfont,
  backgroundcolor=\color{backcolour},   
  commentstyle=\itshape\color{codegreen},
  keywordstyle=\color{codepurple},
  stringstyle=\color{codered},
  numberstyle=\ttfamily\footnotesize\color{darkgray}, 
  breakatwhitespace=false,         
  breaklines=true,                 
  captionpos=b,                    
  keepspaces=true,                 
  numbers=left,                    
  numbersep=5pt,                  
  showspaces=false,                
  showstringspaces=false,
  showtabs=false,                  
  tabsize=2
}

\makeatletter
\g@addto@macro{\UrlBreaks}{\UrlOrds}
\makeatother

\begin{document}
\title{A Chromium-based Memento-aware Web Browser}

\author{Abigail Mabe}
\affiliation{
  \department{Department of Computer Science}
  \institution{Old Dominion University}
  \city{Norfolk}
  \state{VA}
  \postcode{23529}
  \country{USA}
}
\email{amabe002@odu.edu}

\begin{abstract}
Web browsers provide a user-friendly means of navigating the web. Users rely on their web browser to provide information about the websites they are visiting, such as the security state. Browsers also provide a user interface (UI) with visual cues about each tab that is open, including icons for if the tab is playing audio or requires authentication to view. However, current browsers do not differentiate between the live web and the past web. If a user loads an archived webpage, known as a memento, they have to rely on UI elements present within the page itself to inform them that the page they are viewing is not the live web. Additionally, memento-awareness extends beyond recognizing a page that has already been archived. The browser should give users the ability to archive live webpages, essentially creating mementos of webpages they found important as they surf the web. In this report, the process to create a proof-of-concept browser that is aware of mementos is presented. The browser is created by adding on to the implementation of Google’s open source web browser, Chromium. Creating this prototype for a Memento-aware Browser shows that the features implemented fit well into the current Chromium implementation. The user experience is enhanced by adding the memento-awareness, and the changes to the Chromium code base are minimal.
\end{abstract}

\maketitle

\section{Introduction}

\begin{figure*}[ht]
\centering
\includegraphics[width=0.91\textwidth, frame]{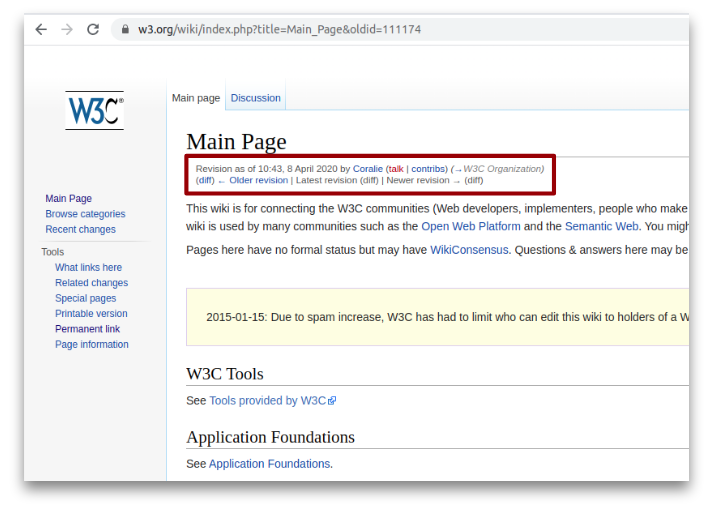}
\caption{Memento of an old revision of a w3.org page.}
\label{fig:w3-example}
\end{figure*}

\begin{figure*}[ht]
\centering
\includegraphics[width=0.91\textwidth, frame]{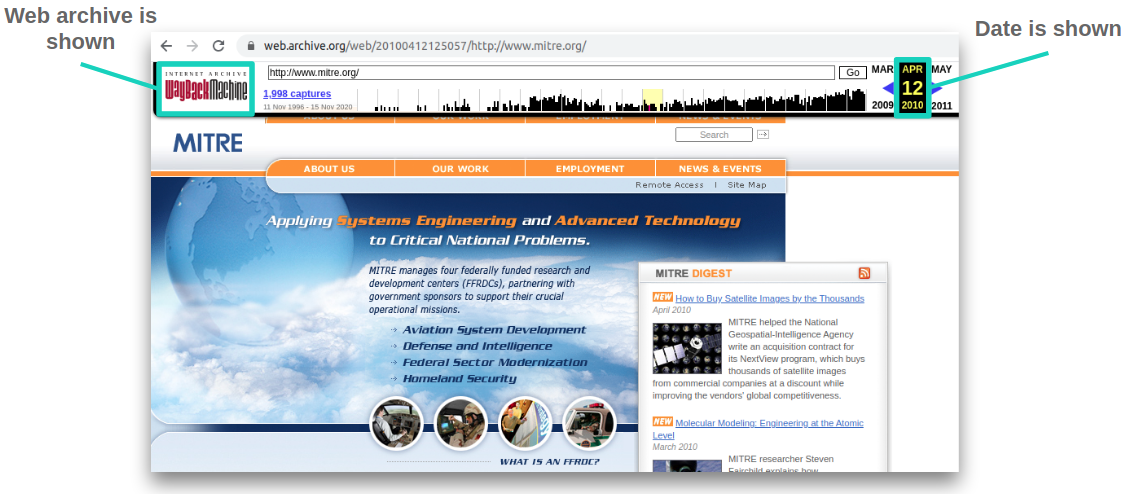}
\caption{Memento via the Internet Archive's Wayback Machine \cite{wayback}. Memento Datetime and web archive information are clearly visible to the user.}
\label{fig:ia-example}
\end{figure*}

The Web is continuously becoming a bigger part of everyday life. The first thing many people do when they want some information is turn to their web browser and type into the search bar. People will search the Web for news, articles, weather reports, or even tutorials if they want to learn a new hobby or skill. We rely on the Web browser to allow us to navigate the Web safely and securely. Web browsers will process information about the webpage as it loads and present us with appropriate user interface cues and messages about the page, keeping us informed as we navigate the Web. For example, browsers can detect if a connection is secure, and we have become used to the browser warning us when a website is insecure or could potentially take our information. The HTTPS secure lock icon is there to quickly tell us that the page we are viewing is safe. Browsers also check for more information about the page other than just security. If a tab is playing music or a video, the browser can add an icon to that tab to let us know that the website displayed by that tab is attempting to play audio. 
The Web is a great resource for information, however it is extremely dynamic. Not only are new webpages being created, but existing ones are changing or being deleted. For these reasons, archiving the Web has become increasingly important. Archiving a webpage produces what is known as a memento \cite{vandesompel2009memento}. Mementos allow you to go back in time on the Web, which is useful for either studying the past Web or retrieving information from a webpage that was changed or deleted. One of the easiest ways you can view mementos is by visiting a web archive, such as the Internet Archive \cite{internet-archive} or Trove \cite{trove}. Mementos can also be found outside of web archives \cite{memento:mediawiki}. For example, \url{w3.org} stores old revisions of their webpages as mementos, which is useful for visiting the page exactly as it was in the past. Figure \ref{fig:w3-example} shows one of these old revisions and circled in red is where you can find the date the page is from, which for this particular memento is April 8th, 2020.
Currently, archived webpages are not recognized by web browsers, meaning that the browser does not react any differently if the page is archived or a part of the live Web. Because of this, the user has to look out for visual cues on the page itself to see if the webpage they are viewing is from the past or current Web. Web archives do a great job of showing the datetime the memento was captured, in addition to displaying other information about the archived page (Figure \ref{fig:ia-example}). However, mementos such as the one shown in Figure \ref{fig:w3-example} may have a datetime visible on the page that is hard to locate, if they do have a datetime visible at all. As archived webpages become increasingly common, this can become an issue and confuse users. A user could even be led to believe that they are on a different page than the one they actually navigated to.

\section{Basics of Modifying Chromium}
\label{section:design}

The Memento-aware Browser was created by adding on to the implementation of Chromium, Google’s open source web browser. This section describes important details about working with the Chromium code base, such as some differences between how it works on different operating systems and how to successfully add on to the implementation. The section also details the proposed features of the Memento-aware Browser and the proposed design that was implemented. Much of the initial design of the features as well as their final implementation was shaped around the current implementation and user experience (UX) of the Chromium browser.

\subsection{Getting Started with the Source Code}
\label{section:working-with-chromium}

To begin working with Chromium, the repository must first be properly downloaded along with the necessary tools. The Chromium source code \cite{chromium-source} is readily available for download. Additionally, Google makes their tools that are required to build and run Chromium, called \lstinline{depot_tools} \cite{depot_tools}, available for download.

\subsection{Downloading and Building Chromium}

Google provides detailed instructions to download, build, and run the Chromium source code for Linux, macOS, and Windows. The Chromium source code also has a GitHub mirror \cite{chromium-github}, however the documentation advises that you do not download and run the code from GitHub. Instead, Google advises that you first download their \lstinline{depot_tools} and run \lstinline{fetch chromium} to download the latest source code for your operating system. Since running \lstinline{fetch chromium} downloads operating system specific code, the same code base will not be able to build and run the Chromium browser on all operating systems, which is why cloning the GitHub mirror is not recommended. 

Before being able to run \lstinline{fetch chromium} and get the source code, you must first download \lstinline{depot_tools} which provides the necessary tools for running Chromium specific commands. Instructions on how to set up the path for \lstinline{depot_tools} varies between operating systems.

Since the Chromium code base varies slightly between operating systems, it is important to note that the Memento-aware Browser was primarily implemented on Linux, specifically Ubuntu 20.04. Additionally, the browser was tested on Windows. Releases for Linux and Windows were generated and are available to download on the Memento-aware Browser GitHub Repository \cite{repo}. The May 2020 Chromium version that was used as a base for the browser did not run on the current version of macOS Catalina, so a macOS version was not created.

\subsection{Editing the Chromium Code}

The majority of the Chromium implementation that was edited for the Memento-aware Browser was in C++. Editing these C++ files makes changes for all operating systems. Occasionally, it is necessary to make operating system specific changes to the code. Listing \ref{lst:os_spec_header_files} shows an example of including different header files depending on the operating system. As you can see in the listing, if the operating system is Windows then \lstinline{base/base_paths_win.h} will be included. If the operating system is macOS or Android, then the header files specific to those operating systems will be included. In addition to including header files depending on the operating system, it is also possible to include different code blocks depending on the operating system. Listing \ref{lst:os_spec_if_statement} shows an if statement executing only if the current operating system is macOS.

% Operating system specific header files
\begin{lstlisting}[language=C++, caption=Include different header files depending on the operating system., label=lst:os_spec_header_files, ]
#if defined(OS_WIN)
#include "base/base_paths_win.h"
#elif defined(OS_MACOSX)
#include "base/base_paths_mac.h"
#elif defined(OS_ANDROID)
#include "base/base_paths_android.h"
#endif
\end{lstlisting}

% Operating system specific code
\begin{lstlisting}[language=C++, caption=Creating operating system specific code., label=lst:os_spec_if_statement, ]
#if defined(OS_MACOSX)

// Operating system specific code here

#endif  // defined(OS_MACOSX)
\end{lstlisting}

\subsection{Debugging the Code}

When running Chromium in debug mode, the easiest way to print debug logs is with \lstinline{DVLOG(0)}. You can further control when debug statements are printed by passing different parameters to the \lstinline{DVLOG()} function as outlined in the Chromium docs \cite{chromium-logging}. Listing \ref{lst:dvlog} shows an example of \lstinline{DVLOG(0)} being used to print a debug statement, and Listing \ref{lst:dvlog-output} shows the output of that line.

% Using DVLOG(0) 
\begin{lstlisting}[language=C++, caption=Using DVLOG(0) for debugging., label=lst:dvlog, ]
DVLOG(0) << "Datetime: " << memento_datetime;
\end{lstlisting}

% Output of DVLOG(0) 
\begin{lstlisting}[language=bash, caption=Output of DVLOG(0)., label=lst:dvlog-output, ]
Datetime: Tue, 05 Mar 2019 09:38:34 GMT
\end{lstlisting}

\subsection{Adding Additional Classes}

\begin{figure*}[ht]
\centering
\includegraphics[width=0.6\textwidth, frame]{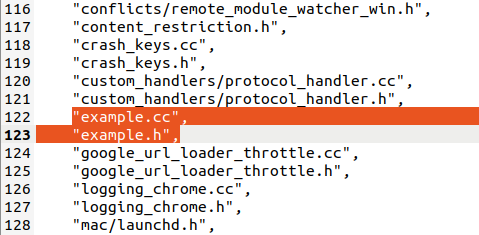}
\caption{Adding a new class to \lstinline{Memento-aware-Browser/src/chrome/common/BUILD.gn}}
\label{fig:build-gn}
\end{figure*}

In order to implement the features of the Memento-aware Browser, it was necessary to not only alter existing C++ classes but also to create entirely new classes. Creating and adding new files to add a new class to the Chromium code is a simple multi-step process. First, you must create the files you need and save them to the appropriate directory. For example, to add the files \lstinline{example.cc} and \lstinline{example.h} to \lstinline{Memento-aware-Browser/src/chrome/common} you must create and save them in the \lstinline{Memento-aware-Browser/src/chrome/common} directory. Next, you must locate the \lstinline{BUILD.gn} file within the \lstinline{Memento-aware-Browser/src/chrome/common} directory. Add the lines \lstinline{"example.cc"} and \lstinline{"example.h"} to the \lstinline{BUILD.gn} as shown in Figure \ref{fig:build-gn}. With the appropriate changes made to the source code, the application can be built with the new classes now included.

\subsection{Adding Additional Icons}

The Chromium browser uses the Skia Graphics Library \cite{skia} as its graphics engine. Any icons rendered by the browser are in the Skia format. The contents of the Skia icon file for the HTTPS secure lock icon are shown in Listing \ref{lst:skia-example}. In the listing, three different icon sizes are shown, where the first icon size is the largest and the last icon size is the smallest. The first is 36 x 16, the second is 24 x 34, and so on. The browser picks which size to use depending on what the icon is being rendered for. One way to generate a new icon to use in the browser is to first create an SVG of the icon and then ``Skiafy" the icon using an online tool. The tool that was used to create the memento icon for the Memento-aware Browser was generated from an SVG using a tool called Skiafy by Evan Stade \cite{skiafy}. The output produced by the Skiafy tool is a great starting point to generate any icon to use in the Chromium browser, however it was necessary to make small changes to the output from the tool in order to get the sizing of the icon correct.

\begin{lstlisting}[float, language=bash, caption=Skia file for the HTTPS secure lock icon., label=lst:skia-example, ]
MOVE_TO, 36, 16,
R_H_LINE_TO, -2,
R_V_LINE_TO, -4,
R_CUBIC_TO, 0, -5.52f, -4.48f, -10, -10, -10,
CUBIC_TO, 18.48f, 2, 14, 6.48f, 14, 12,
R_V_LINE_TO, 4,
R_H_LINE_TO, -2,
R_CUBIC_TO, -2.21f, 0, -4, 1.79f, -4, 4,
R_V_LINE_TO, 20,
R_CUBIC_TO, 0, 2.21f, 1.79f, 4, 4, 4,
R_H_LINE_TO, 24,
R_CUBIC_TO, 2.21f, 0, 4, -1.79f, 4, -4,
V_LINE_TO, 20,
R_CUBIC_TO, 0, -2.21f, -1.79f, -4, -4, -4,
CLOSE,
MOVE_TO, 24, 34,
R_CUBIC_TO, -2.21f, 0, -4, -1.79f, -4, -4,
R_CUBIC_TO, 0, -2.21f, 1.79f, -4, 4, -4,
R_CUBIC_TO, 2.21f, 0, 4, 1.79f, 4, 4,
R_CUBIC_TO, 0, 2.21f, -1.79f, 4, -4, 4,
CLOSE,
[...]
CLOSE
\end{lstlisting}

\section{Memento-Aware Design}

\subsection{Basic Memento Detection}
\label{section:basic-memento-detection}

\begin{figure*}[ht]
\centering
\includegraphics[width=0.8\textwidth, frame]{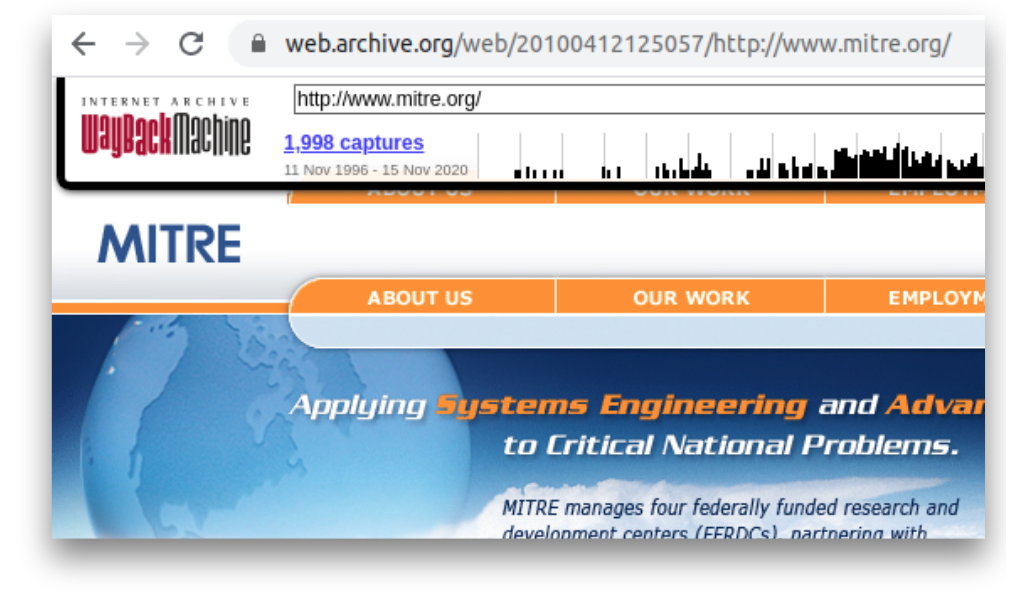}
\caption{Memento of mitre.org in 2010 via the Wayback Machine \cite{wayback}}
\label{fig:wayback-mitre}
\end{figure*}

\begin{figure*}[ht]
\centering
\includegraphics[width=0.8\textwidth, frame]{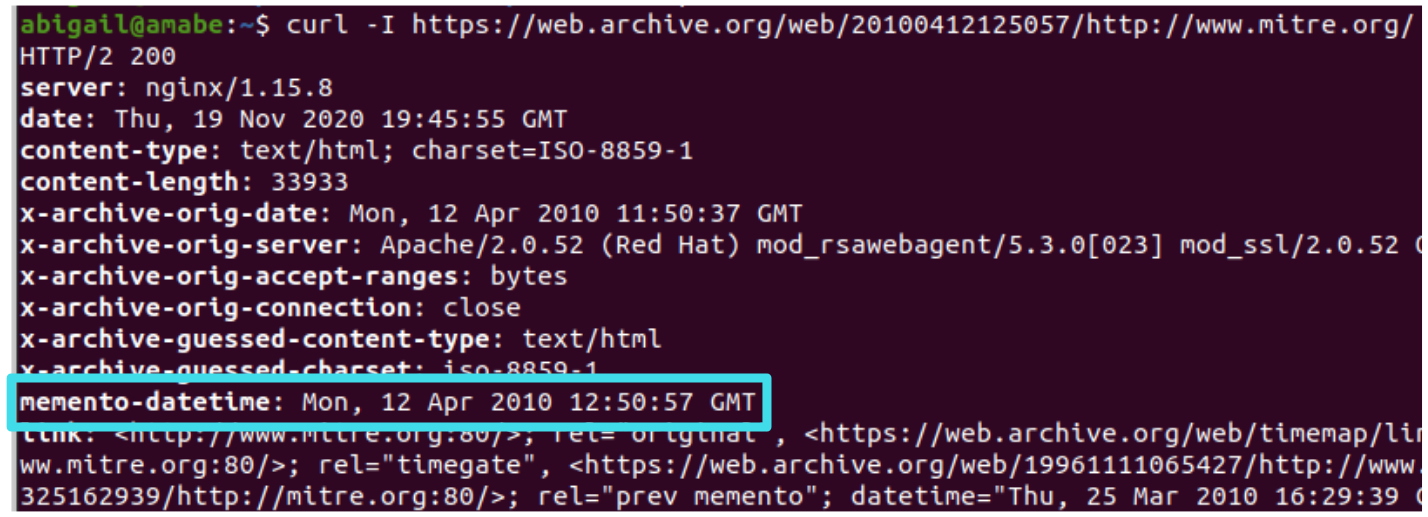}
\caption{Curl request for https://web.archive.org/web/20100412125057/http://www.mitre.org/}
\label{fig:curl-wayback-mitre}
\end{figure*}

The first task in implementing memento detection is to first consider the most basic of possibilities. This would be when the entire root webpage is archived and considered to be a memento. You can easily view this case by looking at an archived page within a web archive. Figure \ref{fig:wayback-mitre} is an example of an archived page from the Internet Archive’s Wayback Machine \cite{wayback}. In this example, we want the browser to classify the visible page as a memento and alert the user that the page is archived and not live. We cannot accomplish this by looking at the URL of the page since there is no defined format for a memento URL. If we look at the URL for an archived page from the Wayback Machine as shown in Figure \ref{fig:wayback-mitre}, we see that the URL contains a datetime as well as the URL of the original live page. However, if we look at a URL for an archived page from Perma.cc \cite{permacc} we see that there is no datetime or original URL. Instead, there is just a hash. This means memento detection is not as simple as just parsing the URL. A better method is to parse the HTTP response headers to find if the webpage is archived since archived webpages return the Memento-Datetime HTTP response header. The Memento-Datetime header indicates that the response contains a representation of a memento where the value of the header is the datetime of the original resource \cite{mementoweb}. The Memento-Datetime header can be observed in Figure \ref{fig:curl-wayback-mitre} where a  simple HEAD request with Curl was made for \href{https://web.archive.org/web/20100412125057/http://www.mitre.org/}{web.archive.org/web/20100412125057/http://www.mitre.org/}.

\subsection{Detecting Embedded Mementos}

\begin{figure*}[ht]
\centering
\includegraphics[width=0.8\textwidth, frame]{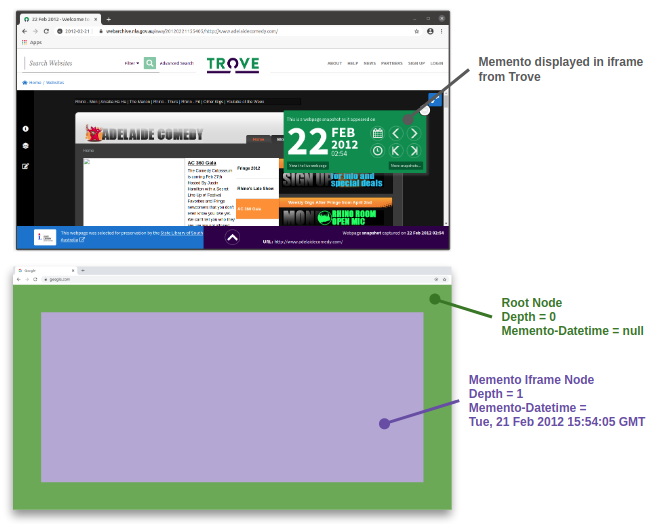}
\caption{Memento displayed within an iframe via Trove}
\label{fig:trove-iframe-memento}
\end{figure*}

\begin{figure*}[ht]
\centering
\includegraphics[width=0.8\textwidth, frame]{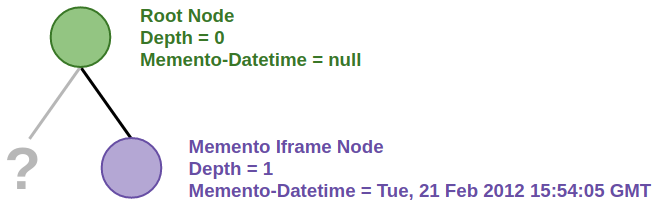}
\caption{Tree structure of a live root webpage from Trove containing an iframe Memento}
\label{fig:iframe-memento-tree}
\end{figure*}

\begin{figure*}[ht]
\centering
\includegraphics[width=0.85\textwidth, frame]{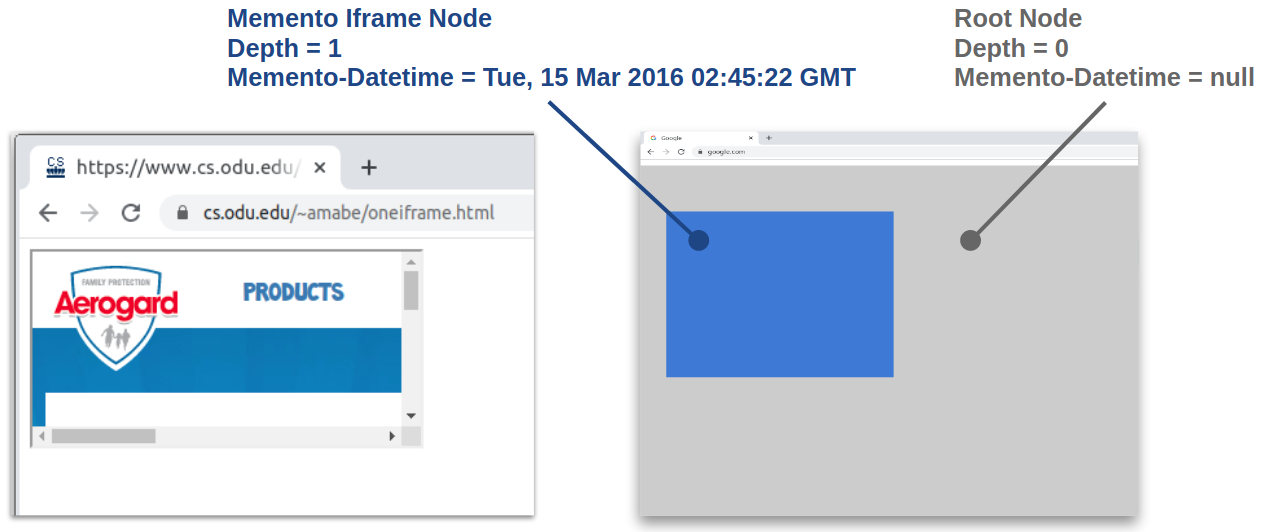}
\caption{Memento displayed within an iframe as an additional page element on a live webpage}
\label{fig:oneiframe}
\end{figure*}

\begin{figure*}[ht]
\centering
\includegraphics[width=0.85\textwidth, frame]{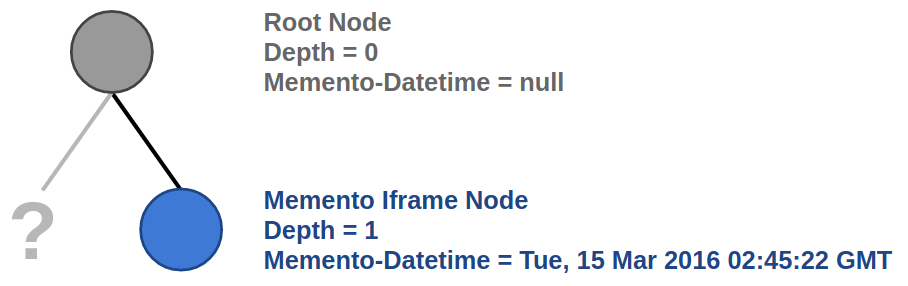}
\caption{Tree structure of a live root webpage containing an iframe Memento}
\label{fig:oneiframe-tree}
\end{figure*}

\begin{figure*}[ht]
\centering
\includegraphics[width=0.85\textwidth, frame]{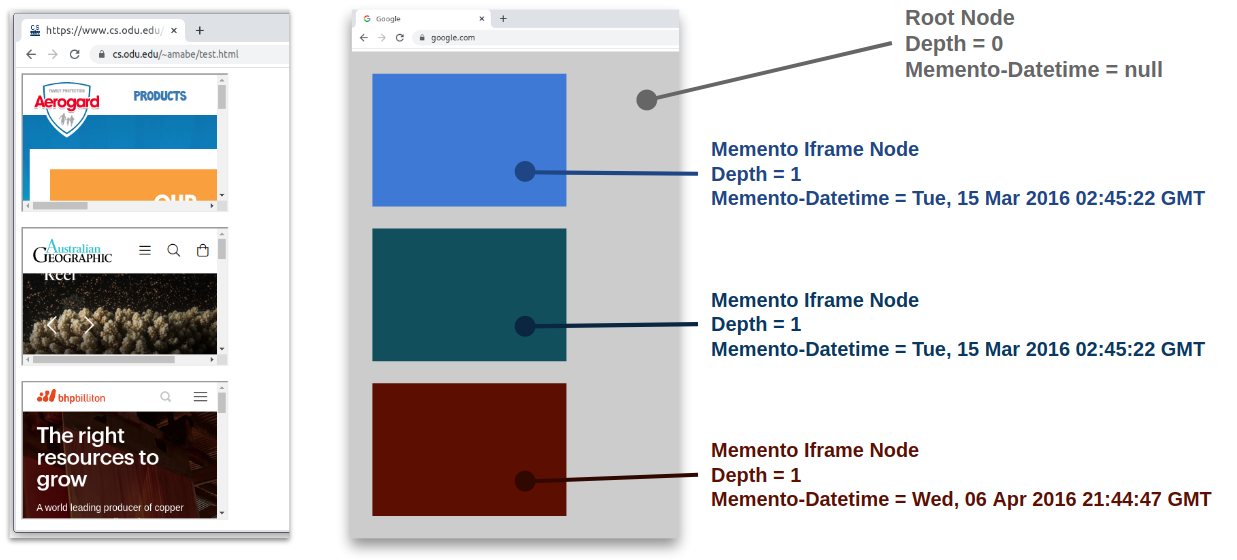}
\caption{Three mementos displayed within iframes as additional page elements on a live webpage}
\label{fig:multiframe}
\end{figure*}

\begin{figure*}[ht]
\centering
\includegraphics[width=0.85\textwidth, frame]{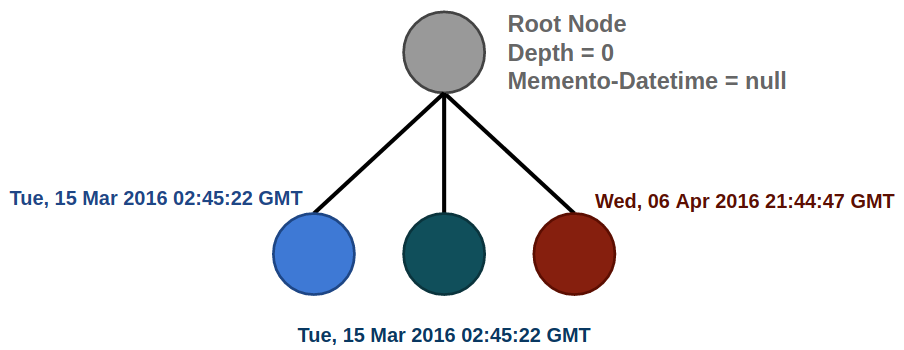}
\caption{Tree structure of a live root webpage containing three iframe Mementos}
\label{fig:multiframe-tree}
\end{figure*}

A memento does not have to be the root webpage. It is possible that a live root webpage could contain one or more embedded archived frames within itself. Most commonly this occurs when the root webpage contains an iframe that loads an archived page. When it comes to these embedded mementos, there are three possibilities:

\begin{itemize}
  \item The first possibility is that it is intended for the iframe memento to make up the whole webpage and the root page should be considered a memento even though the URL of the root page does not return the Memento-Datetime header. This case can be seen in web archives that display the memento within an iframe on the root page, rather than making the memento the root page itself. Examples of such web archives are Trove \cite{trove} and Perma.cc \cite{permacc}. Figure \ref{fig:trove-iframe-memento} illustrates this case by showing a memento displayed in an iframe from Trove. Additionally, the figure shows a diagram of the rendered frames on the page. In the diagram the green portion represents the root webpage and the purple portion represents the embedded iframe. This page structure can be translated to the tree structure as shown in Figure \ref{fig:iframe-memento-tree} where the root webpage is the root node of the tree and the iframe memento is a child node of the root page.
  \item The next possibility is that a live webpage is simply displaying a memento without intending for the memento to make up the whole page. A simple example of this would be when a live page decides to place an iframe memento within their website so their readers can view the content of that memento. To demonstrate this, an example page was created at \url{https://www.cs.odu.edu/~amabe/oneiframe.html}. This is similar to the first possibility in that there is a single memento associated with the current page being viewed. However, the iframe memento is not intended to make up the entire webpage. Instead, it is intended to be an additional element on the page. This possibility is shown in Figure \ref{fig:oneiframe} where the test webpage is shown on the right and on the left is a diagram of the page structure. In the diagram, the gray portion represents the root webpage with no Memento-Datetime header and the blue portion represents the iframe memento. This can also be translated into a tree structure, shown in Figure \ref{fig:oneiframe-tree}. Note that this tree structure is identical to the structure of the memento from Trove shown in Figure \ref{fig:iframe-memento-tree}. As a user, it is easy to differentiate the Trove iframe example from the test webpage example, but from the tree structure the two examples appear the same.
  \item The final possibility, shown in Figure \ref{fig:multiframe}, is when a live webpage has multiple mementos on the page. The context of this is similar to the previous possibility, but instead of a single memento embedded on the currently viewed page there are multiple embedded mementos. An example webpage was created to demonstrate this and can be viewed at \url{https://www.cs.odu.edu/~amabe/test.html}. Again, this can be thought of as a tree structure as shown in Figure \ref{fig:multiframe-tree} where the root node (root webpage) has 3 child nodes (iframe mementos).
\end{itemize}

Each of these possibilities should be detected by the browser in order to present the user with the correct message about the content they are viewing. Considering the first case, if the user were to view an archived page through Trove or Perma.cc the browser should detect the datetime from the archived content in the iframe and detect that the iframe is formatted to make up the entire page. When these things are detected, the browser should then tell the user that the page they are viewing is an archived page from a past date. In the second possibility, the browser should detect the archived content within the page but also detect that the embedded iframe is an additional element on the live page. The user should be able to tell through the native UI elements in the browser that the page they are viewing contains archived content and they should also be able to read the datetime that the archived content is from. In the third and final case, the browser should behave similarly as to when the second possibility occurs. The only difference is that the browser should list all datetimes from all archived content on the page rather than just the one datetime. 

\subsection{Embedded Elements Within Mementos}
\label{section:embedded-elements-within-mementos}

As covered in the previous section, a root webpage can display a number of embedded elements. It is possible that the root webpage is an archived page, or an embedded element on the root page is archived. The previous cases covered were about live webpages displaying archived content. The next possible cases to cover involve an archived page or embedded archived element displaying its own embedded elements that may or may not also be archived. The archived page displaying embedded content could be the root page or an embedded element itself. Either way, an archived page or element can display its own embedded elements and they may or may not also be archived which changes how the browser should react to the archived content.

\begin{figure*}[ht]
\centering
\includegraphics[width=0.8\textwidth, frame]{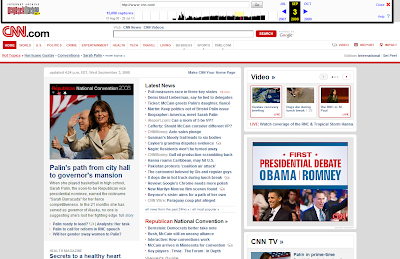}
\caption{``Zombie" memento from 2008 that is displaying an advertisement from 2012 \cite{zombies}}
\label{fig:best_one}
\end{figure*}

\begin{itemize}
  \item The first case is that an archived frame (could be the root page or an element on the root page) contains one or more embedded elements that are displaying content from the live web. This is a rare yet undesirable occurrence since the archived webpage is intended to present content from the past, yet there is a portion or portions of the page that can display content from the live web inside the memento.  When this occurs, it is not necessarily obvious to the user that the live web is leaking into the archived page. Such an occurrence could be called a ``zombie" \cite{zombies}. A common example of this occurring is when the archived page contains Javascript or an embedded element that pulls content from the live web. Figure \ref{fig:best_one} illustrates this. In the figure, it can be seen that the memento is from 2008 but there is an advertisement from 2012.
  \item The second case is that an archived page contains one or more embedded elements that are displaying additional archived frames, each with their own datetime. This is similar to the first case in that the memento is displaying embedded content, but the content is not from the live web. Typically when this case occurs it is not an issue as the embedded frames that are displayed within the memento were archived with the memento and do not allow the live web to leak in.
\end{itemize}

In these cases, whether the archived frame that is displaying embedded content is the root page or an embedded element itself, the browser should parse response headers for any content within the archived frame. If Memento-Datetime headers are found, the user should be alerted that frames with different datetimes exist within the page. If frames within the archived content do not return the Memento-Datetime header, the user should be alerted that the archive content they are viewing contains the live web.

\subsection{Bookmark as Archive}

\begin{figure*}[ht]
\centering
\includegraphics[width=0.8\textwidth, frame]{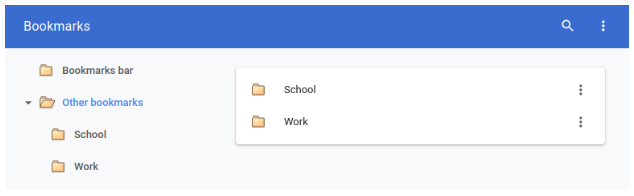}
\caption{Bookmark folders within the Chromium bookmark manager}
\label{fig:bookmark-folders}
\end{figure*}

The concept of a Memento-aware browser goes beyond the browser detecting archived content. In addition to making users aware of any archived content they are viewing, users should also be able to easily archive any live content that they want to save. Essentially, when the user saves a bookmark they should have the option to also archive it so that when they go to their bookmarks, they have the option to go to the live web or view the archived version in a public web archive \cite{weigle-wadl18}. To design a feature that allows users to submit a live webpage to be archived, it is best to first think of what a user currently does to save a webpage. The common way that a user saves a webpage for viewing at a later date is by bookmarking the page through their browser. The bookmark functionality saves a link to the page within the browser settings and allows the user to revisit that link in the location they saved it. For example, saving a bookmark to the bookmarks bar places a link to the page in a readily accessible location in the browser UI. The user also has the option to save the bookmark to a folder. The user can create a bookmark folder named “School” and save any relevant webpages to their “School” bookmark folder as shown in Figure \ref{fig:bookmark-folders}. This bookmarking functionality is a great way to save and organize webpages for later viewing. However, bookmarking only saves a link to the page. If the page were to be taken down, the bookmark link would lead to a 404. Additionally, if the content on the webpage were to change, the bookmark link would still lead to the correct page but the content that the user originally saw and wanted to save would no longer be present. In many cases we can assume that the page the user wanted to save contained important information, which is why they wanted to bookmark it. In the case of 404s and changed content, that important information that the user wanted to save is, at best, temporarily lost and the information on the page may be available again at a later date or is now available elsewhere. In the worst case scenario that content is permanently lost.

A way to make these bookmarked pages constant is through the use of public web archives. When the user wants to bookmark a webpage, the browser should also allow them to submit the webpage to a public web archive. This way, the browser bookmarking functionality would work as normal where a link to the webpage would be placed in the proper bookmark folder. However in addition to this, the browser would also submit the webpage to the user selected public web archive. The browser would create an additional bookmark folder for this archive and place the link to the webpage in the archive within this new folder. The end result of this design is that the user has their traditional bookmark system, but if they encounter a 404 or a changed webpage they can access their archived bookmarks and visit the page as it was the day that they originally wanted to save it.

\section{Implementation}

The design and features described in the previous section were implemented in a way that allowed the features to be added on to the native implementation of the browser with minimal changes to the native code. The majority of the code for the Chromium web browser is C++ so most of the implementation was done in C++. The Chromium browser does also have HTML5, Javascript, CSS3, and Python files that make up things such as settings pages and the bookmark manager for the browser. Some of these files were edited as needed. To build and run the browser, Google provides their own \lstinline{depot_tools} and the Chromium browser uses the Ninja build system, so the instructions listed by The Chromium Projects \cite{install-dependencies} were followed in order to build and run the browser. 

The primary environment used to develop the Memento-aware Browser has the following specifications:

\begin{itemize}
	\item Ubuntu 20.04.1 LTS 64-bit operating system
	\item 16 GB memory
	\item AMD Ryzen 5 3600 3.6 GHz 6-Core Processor
	\item 1TB solid state drive
	\item Python 3.8.5
	\item HTML5 and CSS3 for webpages standard to the browser (settings, profile, etc.)
\end{itemize}

Additionally, the Memento-aware Browser was tested on a Windows system with the following specifications:

\begin{itemize}
	\item Windows 10 Home 64-bit operating system
	\item 16 GB memory
	\item Intel(R) Core(TM) i7-8550U CPU
	\item 1TB solid state drive
	\item Python 3.8.0
	\item HTML5 and CSS3 for webpages standard to the browser (settings, profile, etc.)
\end{itemize}

The implemented features were primarily done in C++ code that works the same between operating systems. The primary difference between operating systems that should be noted is required dependencies. The GitHub repository for the Memento-aware Browser \cite{repo} can be built and run on Linux and Windows as is. With the proper dependencies and up to date Chromium version as a base, the code could potentially also be run on macOS.

\subsection{Chromium Page Structure}
\label{section:chromium-page-structure}

\begin{figure*}[ht]
\centering
\includegraphics[width=0.6\textwidth, frame]{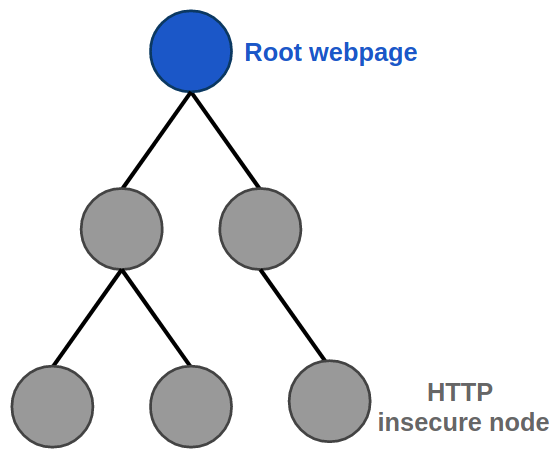}
\caption{Chromium treats rendered frames as nodes in a tree with the root node being the root page. An insecure HTTP node can cause an insecure security state for the page.}
\label{fig:chromium-page-structure}
\end{figure*}

The implementation for the Memento-aware Browser is built off of the Chromium browser page structure and the state of the page security information. The Chromium browser considers a single tab as an entry where the current active tab is known as an Active Entry. The Active Entry contains information about the page and all resources that make up that page. A part of the information about the page contained within the Active Entry information is the Security State. The Security State contains security information about the page such as whether the protocol of the root webpage is secure or insecure, and if the webpage contains any content that is insecure. The Active Entry also consists of a tree of rendered frames that exist on the page. The root node of the tree is the root webpage, and child nodes of the root node are rendered frames on the root page. This tree structure is illustrated in Figure \ref{fig:chromium-page-structure}. Also shown in the figure is a single insecure HTTP node. Since this one child node on the tree is considered insecure, this makes the entire Active Entry considered insecure since it contains a mix of secure and insecure content.

To include archived page information in the Active Entry, the Security State has been extended to contain Memento-Datetime response header information found while parsing response headers.

\subsection{Detecting the Memento-Datetime Header}
\label{section:detecting-memento-header}

To implement the memento information portion of the Security State, the Memento-Datetime response header first needs to be detected. Additionally, as described in the Section \ref{section:design}, there are a variety of ways that archived content needs to be detected and a variety of messages to present to the user based on the presence of archived content. Properly detecting archived content begins with parsing any received response headers when a page navigation is requested to find the Memento-Datetime header. When this header is found, the datetime also needs to be extracted and associated with the resource that returned that particular memento datetime. Within the Chromium browser, response headers from all resources are parsed within the URLLoader class. The URLLoader class constructs a URLResponseHead object to hold response information for that particular resource. A URLResponseHead object holds information about the resource such as the URL itself and the \lstinline{content_length} of the resource. To implement memento detection, the URLResponseHead object has been extended to hold a \lstinline{memento_datetime} string that is an empty string by default. If the Memento-Datetime header is found when the response headers are parsed, then the \lstinline{memento_datetime} member of the URLResponseHead object is set to the value of the Memento-Datetime header (Listing \ref{lst:urlresponsehead-mementodatetime}). 

\begin{lstlisting}[language=C++, caption=Finding the Memento-Datetime header and setting the \lstinline{memento_datetime} of the URLResponseHead object., label=lst:urlresponsehead-mementodatetime, ]
// Check for Memento-Datetime header.
if (response->headers->HasHeader("Memento-Datetime") || 
  response->headers->HasHeader("memento_datetime")) {

	response->memento_info = true;
	response->memento_datetime = response->headers->GetMementoDatetime();

}
\end{lstlisting}

The URLResponseHead object can then be accessed by the currently committing NavigationRequest. Listing \ref{lst:navigationrequest-mementodatetime} shows the member function \lstinline{GetMementoDatetime()} of the NavigationRequest class where the URLResponseHead object returns the \lstinline{memento_datetime}.

\begin{lstlisting}[language=C++, caption=Accessing the Memento-Datetime value from the URLResponseHead., label=lst:navigationrequest-mementodatetime, ]
std::string NavigationRequest::GetMementoDatetime() {
	if(response()) {
		return response()->memento_datetime;
	}
	return "";
}
\end{lstlisting}

\subsection{Root Page}

With parsing response headers for the Memento-Datetime header implemented within the \lstinline{URLLoader} class, root page memento detection could be further implemented. When the user navigates to a webpage, a \lstinline{NavigationRequest} is committed. The \lstinline{Navigator} class initiates the navigation and utilizes the \lstinline{NavigationController} class to control the currently rendered frame (\lstinline{RenderFrameHostImpl}) and the navigation entry (\lstinline{NavigationEntry}). Since the \lstinline{URLResponseHead} object for the root page contains the \lstinline{memento_datetime} member variable, this datetime string can be accessed when a \lstinline{NavigationRequest} is committed. Using the \lstinline{NavigationRequest::GetMementoDatetime()} function, the currently rendering frame member variable, \lstinline{params->memento_datetime}, is set to the \lstinline{memento_datetime} of the currently committing navigation request (Listing \ref{lst:renderframehostimpl-mementodatetime}).

\begin{lstlisting}[float, language=C++, caption=Within \lstinline{render_frame_host_impl.cc} set the \lstinline{memento_datetime} of the current rendered frame to the \lstinline{memento_datetime} of the root page resource., label=lst:renderframehostimpl-mementodatetime, ]
void RenderFrameHostImpl::DidCommitPerNavigationMojoInterfaceNavigation(...) {
  ...

  params->memento_datetime = committing_navigation_request->GetMementoDatetime();

  ...
}
\end{lstlisting}

With the rendering frame's \lstinline{memento_datetime} parameter properly set to the value that was stored in the \lstinline{URLResponseHead}, the \lstinline{memento_datetime} can then be set for the entire \lstinline{NavigationEntry} within the \lstinline{NavigationController} class (Listing \ref{lst:activeentry-mementodatetime}).

\begin{lstlisting}[float, language=C++, caption=Set the \lstinline{memento_datetime} of the \lstinline{NavigationRequest} object to the value stored in \lstinline{params.memento_datetime}., label=lst:activeentry-mementodatetime, ]
if (params.memento_datetime != "") {
  active_entry->SetMementoDatetime(params.memento_datetime);
  active_entry->SetMementoInfo(true);
}
\end{lstlisting}

\begin{figure*}[ht]
\centering
\includegraphics[width=0.8\textwidth, frame]{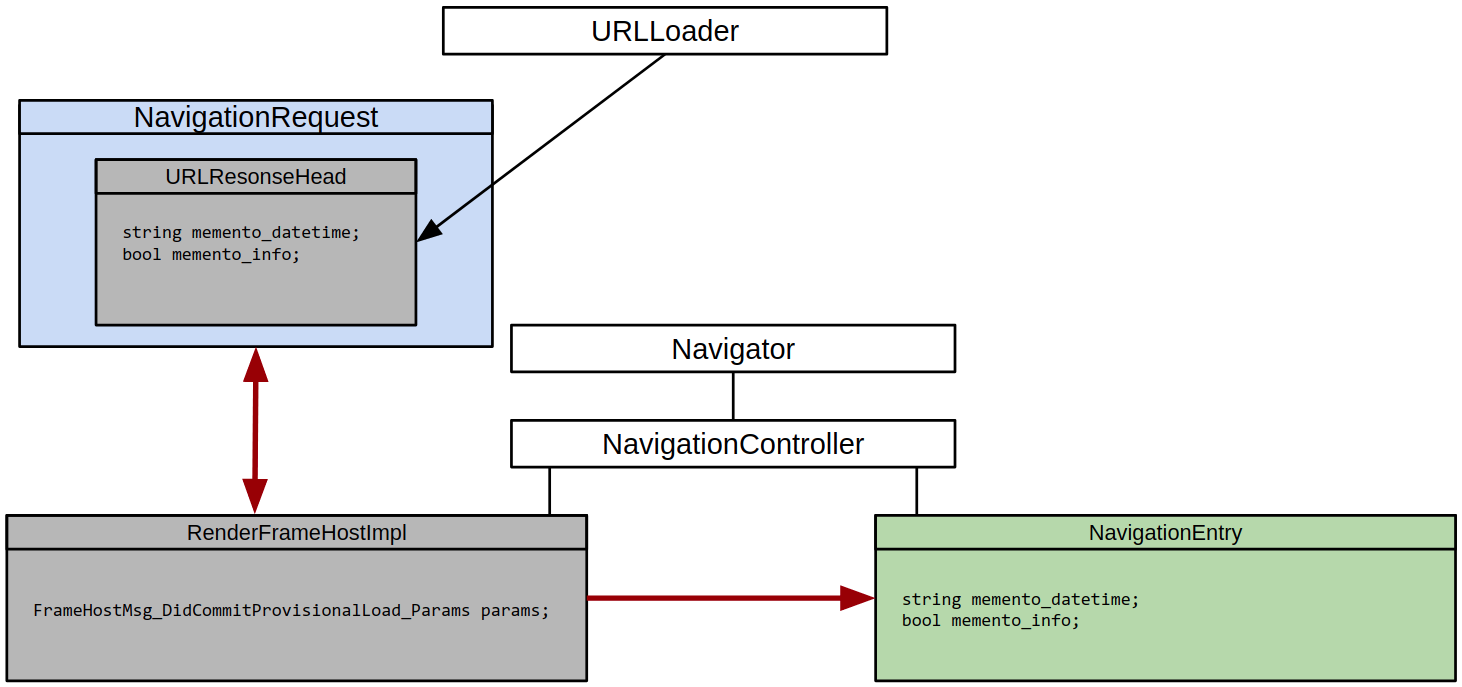}
\caption{The sequence of classes involved in loading a webpage. These are the main classes that were edited to implement rootpage memento detection.}
\label{fig:rootpage-loading-backend}
\end{figure*}

The \lstinline{NavigationEntry} object that now holds the proper memento information variables for the page acts as the bridge between the back-end and front-end of the browser as the page loads. Figure \ref{fig:rootpage-loading-backend} illustrates the previously described process of detecting the Memento-Datetime header when the URL is loaded and bringing the value up to the where it can be accessed by the front-end. In the figure, when the user loads a webpage and a \lstinline{NavigationRequest} is initiated, it creates a \lstinline{URLResponseHead} object that contains the memento information member variables, along with other necessary variables for the loaded URL. The \lstinline{URLLoader} class loads the URL for the root page and parses the response headers. If the Memento-Datetime header is found, the values of the memento information member variables of the \lstinline{URLResponseHead} are set accordingly. Next, the \lstinline{Navigator} can begin processing the navigation to the webpage. The \lstinline{Navigator} uses a \lstinline{NavigationController} to control the \lstinline{RenderFrameHostImpl} object (the currently rendering frame) and the \lstinline{NavigationEntry} object (the entry for the navigation to that particular webpage). The parameters of the rendering frame are set based off of the \lstinline{URLResponseHead} object for that particular frame and its respective resources. There is a single main \lstinline{NavigationEntry} object being used for a webpage navigation, but the currently rendering frame changes depending on what frame within the page is currently loading. Because of this, the \lstinline{NavigationEntry} variables are updated as the currently rendering frame parameters are updated. Thus, the \lstinline{NavigationEntry} contains information regarding all rendered frames within that navigation once loading is fully complete.

The front-end interacts with the \lstinline{NavigationEntry} through the \lstinline{ContentUtils} class. Within \lstinline{content_utils.cc}, various parameters that make up the \lstinline{VisibleSecurityState} are set according to the values within the \lstinline{NavigationEntry} (Listing \ref{lst:rootpage-contentutils}).

\begin{lstlisting}[float, language=C++, caption=Set the \lstinline{VisibleSecurityState} according to the values that were set for the current \lstinline{NavigationEntry} within \lstinline{content_utils.cc}., label=lst:rootpage-contentutils, ]
  // Flag for if the current NavigationEntry is an error page
  state->is_error_page = entry->GetPageType() == content::PAGE_TYPE_ERROR;
  
  // String that hold the Memento-Datetime header value of the NavigationEntry
  state->memento_datetime = entry->GetMementoDatetime();

  // Flag for if the current NavigationEntry is considered a Memento
  state->memento_info = entry->GetMementoInfo();
\end{lstlisting}

With the \lstinline{VisibleSecurityState} for the webpage set, the \lstinline{PageInfo} class can then access the \lstinline{VisibleSecurityState} object and use its parameters to create an \lstinline{IdentityInfo} object for the page. The \lstinline{IdentityInfo} object will then be used to set the values within \lstinline{PageInfoUI} to construct the user interface according to the various security attributes within the \lstinline{VisibleSecurityState} (Listings \ref{lst:rootpage-pageinfo} and \ref{lst:rootpage-pageinfoui}).

\begin{lstlisting}[language=C++, caption=Create an \lstinline{IdentityInfo} object and set the memento information members within \lstinline{page_info.cc}., label=lst:rootpage-pageinfo, ]
  PageInfoUI::IdentityInfo info;

  info.memento_status = memento_status_;
  info.memento_datetime = memento_datetime_;
\end{lstlisting}

\begin{lstlisting}[float, language=C++, caption=Construct the user interface according to the values passed from the \lstinline{IdentityInfo}., label=lst:rootpage-pageinfoui, ]
  if (memento_status && memento_datetime != "") {

    security_description->memento_summary = l10n_util::GetStringUTF16(IDS_PAGE_INFO_MEMENTO_SUMMARY);

    std::string datetime_string = "The page displayed is a memento captured on " + memento_datetime;

    security_description->memento_info = base::UTF8ToUTF16(datetime_string);

  }
\end{lstlisting}

\begin{figure*}[ht]
\centering
\includegraphics[width=0.86\textwidth, frame]{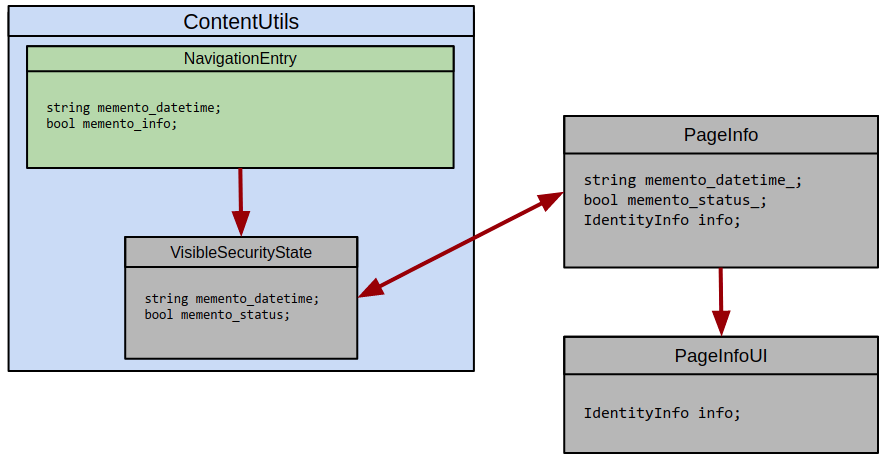}
\caption{The sequence of classes involved in constructing the browser UI for a loaded webpage. These are the main classes that were edited to implement the UI for rootpage memento detection.}
\label{fig:rootpage-loading-frontend}
\end{figure*}

This process of the front-end accessing the \lstinline{NavigationEntry} object to construct the final set of UI elements for the page is illustrated in Figure \ref{fig:rootpage-loading-frontend}. The \lstinline{ContentUtils} class controls the use of the \lstinline{NavigationEntry} object to set the \lstinline{VisibleSecurityState} of the page. \lstinline{ContentUtils} is continuously updating as pieces of the webpage load and the \lstinline{NavigationEntry} updates with new information. The \lstinline{PageInfo} class pulls information from the \lstinline{VisibleSecurityState} and constructs an \lstinline{IdentityInfo} object to be passed to \lstinline{PageInfoUI} where the parameters that make up the front-end of the browser are set accordingly. This is where any messages presented to the user are constructed, such as the "Connection Secure" text or the "Your connection to this site is not secure" message. For the implementation of root page memento detection in the Memento-aware Browser, the message "The page displayed is a memento captured on <datetime>" was added to the list of possible messages where the datetime from the response headers would be added on to the message.

\begin{comment}
if the user should be presented with the HTTPS secure lock icon or the "Not secure" warning. For the Memento-aware Browser implementation, this is also where it is decided if the user should be presented with a memento icon that alerts them that the webpage they are viewing is archived or contains archived content.
\end{comment}

\subsection{Displaying the Memento Icon}

As of this point in the implementation, the front-end of the browser is aware of whether or not the root webpage is an archived page and also the datetime the page was archived, if applicable. The browser can also set the string variable containing the appropriate message about the page being archived. Now the user needs to be presented with the memento icon to be alerted that the page they are viewing is an archived page. This icon should also be clickable so that the user can open up an additional menu providing more details about the archived page they are viewing. To do this, an additional location icon was added to the \lstinline{LocationBarView} as shown in Listing \ref{lst:location_icon_memento} where \lstinline{location_icon_view} is the standard icon that can show the HTTPS secure lock icon and \lstinline{location_icon_memento} is the new memento icon.

\begin{minipage}{\linewidth}
\begin{lstlisting}[language=C++, caption=Add an additional icon to the location bar within \lstinline{location_bar_view.cc}., label=lst:location_icon_memento, ]
  auto location_icon_view =
      std::make_unique<LocationIconView>(font_list, this, this);

  auto location_icon_memento =
      std::make_unique<LocationIconView>(font_list, this, this);
\end{lstlisting}
\end{minipage}

With this extra icon added, there needed to be a way within the icon class to differentiate which is the original icon and which is the memento icon. A boolean flag was added to the \lstinline{LocationIconView} class implementation to tell if that particular icon is the memento icon. By default, this flag is set to false to signify that the icon is the standard icon (Listing \ref{lst:location_icon_flag}). This flag will be used throughout the \lstinline{LocationIconView} class to set various attributes for the icon. For example, if the root webpage is an archived webpage then the datetime for the root page should be displayed next to the memento icon in \lstinline{YYYY-MM-DD} format. This means that if \lstinline{is_memento_icon_} is set to true, then the \lstinline{LocationIconView::ShouldShowText()} function should always return true since the icon should be displaying text. Additionally, the icon resource being displayed by the \lstinline{LocationIconView} should always be set to the memento icon resource since the standard icon is already displaying other necessary information such as the HTTPS secure lock icon.

\begin{lstlisting}[language=C++, caption=Flag for whether or not the \lstinline{LocationIconView} object is the memento icon., label=lst:location_icon_flag, ]
  // Whether the icon is the original icon
  // or the new memento icon
  bool is_memento_icon_ = false;
\end{lstlisting}

Now the memento icon exists within the implementation but does not appear within the location bar by default. The memento icon should only be displayed if the webpage was determine to be a memento. In order for the memento icon \lstinline{LocationIconView} object to determine if it should appear in the location bar, it needs to be able to access the \lstinline{VisibleSecurityState}. To do this, the \lstinline{LocationIconView} class can use the \lstinline{LocationBarModelImpl} to call an \lstinline{IsMemento()} function (Listing \ref{lst:ismemento}).

\begin{lstlisting}[float, language=C++, caption=The memento icon needs to reach the \lstinline{VisibleSecurityState} through the \lstinline{LocationBarModelImpl} to determine if it should be visible in the location bar., label=lst:ismemento, ]
  bool LocationIconView::ShouldShowMementoInfo() const {
	  return delegate_->GetLocationBarModel()->IsMemento();
  }
\end{lstlisting}

The \lstinline{LocationBarModelImpl} class has access to the security information (\lstinline{SecurityState}) that was set in the previous section and can return the \lstinline{memento_status} (Listing \ref{lst:ismemento-status}).

\begin{lstlisting}[float, language=C++, caption=Return the \lstinline{memento_status} stored within the \lstinline{VisibleSecurityState}., label=lst:ismemento-status, ]
	bool LocationBarModelImpl::IsMemento() const {
	  std::unique_ptr<security_state::VisibleSecurityState>
	          visible_security_state = delegate_->GetVisibleSecurityState();

	  return visible_security_state->memento_status;
	}
\end{lstlisting}

\begin{figure*}[ht]
\centering
\includegraphics[width=0.86\textwidth, frame]{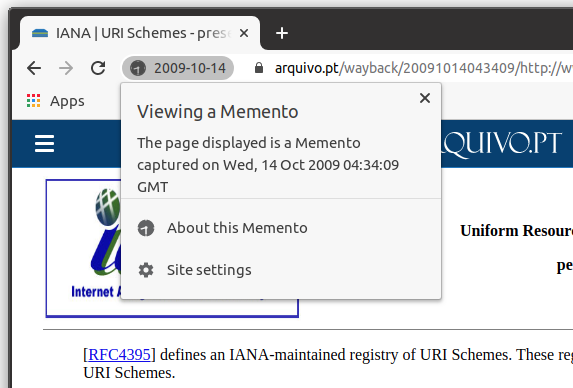}
\caption{The memento icon, datetime in \lstinline{YYYY-MM-DD} format, and popup with additional information for rootpage memento detection.}
\label{fig:rootpage-memento-UI}
\end{figure*}

With the \lstinline{VisibleSecurityState} information being used to determine if the memento icon should become visible, the root page memento detection and user interface is complete. As shown in Figure \ref{fig:rootpage-memento-UI}, when the user loads an archived webpage such as a memento from the Portuguese Web Archive, arquivo.pt \cite{portugeuse}, the browser is able to detect the Memento-Datetime HTTP response header and update the status of the page accordingly, making the memento icon and the datetime of the webpage easily visible for the user.

\subsection{Iframe Elements}
\label{section:iframe-elements-section}

The previous section described detecting when the root webpage returns the Memento-Datetime HTTP response header. As described in the Design section, mementos can appear as embedded elements, not just as the root webpage. There are three possibilities of this occurring to account for in the implementation:

\begin{itemize}
	\item A memento is being displayed as an embedded element and is intended to make up the whole webpage. The best example of this is how the archives Trove and Perma.cc display Mementos.
	\item A live webpage is displaying a memento within an embedded element as an additional part of the page, but the whole webpage is not intended to be considered a memento.
	\item A live webpage is displaying multiple mementos within embedded elements as additional parts of the page, but the while webpage is not intended to be considered a memento.
\end{itemize}

The first case of detecting mementos such as the ones displayed by Trove and Perma.cc has the same end goal as detecting archived root webpages. The Memento-Datetime HTTP response header should be detected by the browser and the user should be presented with the memento icon and the datetime the page was archived. To implement this detection, properties of the root page detection implementation were altered or added on to. For example, as described in the Root Page section, the \lstinline{NavigationEntry} object updates with new information as each frame on the webpage is rendered. However, with the implementation in that section only the root page Memento-Datetime header is considered as a page parameter and passed to the \lstinline{NavigationEntry}. All resources already pass through the \lstinline{URLLoader} class and have any discovered Memento-Datetime header values associated with their corresponding \lstinline{URLResponseHead} object, but these are not immediately passed on to the page parameters since they involve more processing and consideration than the root page datetime. These datetimes associated with rendered frames on the page need to be considered as the entire webpage loads. In the case of a memento via Trove, the datetime from the iframe source should be associated with the entire webpage. Meaning that the \lstinline{VisibleSecurityState} will hold the datetime of the iframe source and the UI will appear the same as if the root page were a memento. 

\begin{figure*}[ht]
\centering
\includegraphics[width=0.85\textwidth, frame]{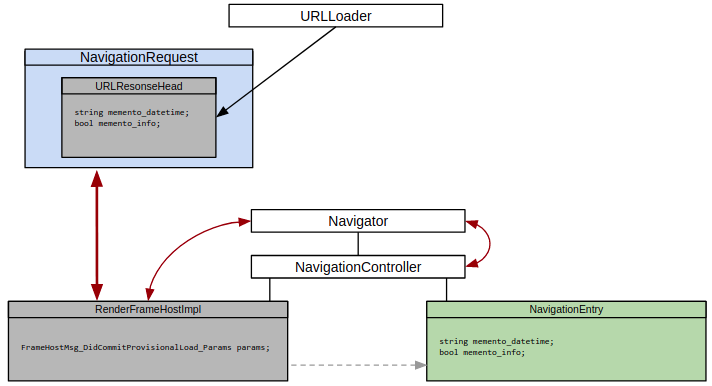}
\caption{The loading sequence for embedded page elements that return the Memento-Datetime header.}
\label{fig:iframe-memento-sequence}
\end{figure*}

The implementation for this is shown in Figure \ref{fig:iframe-memento-sequence}, where the path of the datetime value of frames within the root page is different from the path of the datetime value of the root page itself. The dotted gray line represents where the root page datetime would be passed from the root page \lstinline{RenderFrameHostImpl} object to the \lstinline{NavigationEntry}. For a datetime associated with a frame within the root page, the datetime cannot be immediately passed to the \lstinline{NavigationEntry} because it may or may not be associated with the root page. Instead, the datetime gets passed back up to the \lstinline{Navigator} and then down again to the \lstinline{NavigationController} where it can be determined if it should be taken as the datetime for the entire \lstinline{NavigationEntry}. If the \lstinline{NavigationController} determines that the iframe source is intended to act as the root webpage, the datetime is set as the datetime for the entire page. This is determined by the loading sequence of the nodes on the page. The root page processes last since all resources need to load first, meaning iframes will first be processed as subframes. If the iframe is processed a second time as the root page is processed, then the iframe is considered as a main piece of the root page and not just an element on the page. The \lstinline{switch} used to determine this is shown in Listing \ref{lst:processing-switch}. If the navigation type is \lstinline{NAVIGATION_TYPE_AUTO_SUBFRAME} then the element is an iframe. For iframe mementos that are to be considered as the root page, this \lstinline{switch} case is met the first time the frame is processed. The second time it is processed, the iframe is processed along with the root page. Essentially, if the datetime that was found was the only datetime and the embedded element was processed a second time with the root page, then its datetime will be considered as the datetime for the whole page.

\begin{lstlisting}[float, language=C++, caption=Switch statement that directs the processing of different navigation types where the processing of the \lstinline{AUTO_SUBFRAME} type is for iframe elements and used in the memento Detction implementation., label=lst:processing-switch, ]
	switch (details->type) {
    case NAVIGATION_TYPE_NEW_PAGE:
      ...
    case NAVIGATION_TYPE_EXISTING_PAGE:
      ...
    case NAVIGATION_TYPE_SAME_PAGE:
      ...
    case NAVIGATION_TYPE_NEW_SUBFRAME:
      ...
    case NAVIGATION_TYPE_AUTO_SUBFRAME:

      // iframes processed here

    case NAVIGATION_TYPE_NAV_IGNORE:

      // If a pending navigation was in progress, this canceled it.  We should
      // discard it and make sure it is removed from the URL bar.  After that,
      // there is nothing we can do with this navigation, so we just return to
      // the caller that nothing has happened.
      if (pending_entry_)
        DiscardNonCommittedEntries();
      return false;

    case NAVIGATION_TYPE_UNKNOWN:
      NOTREACHED();
      break;
  }
\end{lstlisting}

The fact that embedded elements are processed in this \lstinline{switch} case means we can collect any datetimes associated with elements into a list, completing the implementation for the other two possibilities for iframe mementos (Listing \ref{lst:collect-memento-dates}). Doing this allows a list of datetimes that exist on the page to be stored in the \lstinline{NavigationEntry}, meaning that the \lstinline{NavigationEntry} class now has a total of 3 variables regarding the memento information for the whole page.

\begin{lstlisting}[float, language=C++, caption=For the \lstinline{AUTO_SUBFRAME} case collect any datetimes returned by embedded elements., label=lst:collect-memento-dates, ]
    case NAVIGATION_TYPE_AUTO_SUBFRAME:
      if (!RendererDidNavigateAutoSubframe(rfh, params, navigation_request)) {

        NavigationEntry* visible_entry = GetVisibleEntry();

        if (datetime != "" &&
            root != frame_tree_node && 
            frame_tree_node->depth() < 2) {

          root->AddMementoDate(datetime);
          visible_entry->SetMementoDates(root->GetMementoDates());
        }

        visible_entry->SetIterations(iterations);
        visible_entry->SetIsMixedMementoLiveWeb(mixed_memento_live_web);

        // We don't send a notification about auto-subframe PageState during
        // UpdateStateForFrame, since it looks like nothing has changed.  Send
        // it here at commit time instead.
        NotifyEntryChanged(GetLastCommittedEntry());
        return false;
      }
      break;
\end{lstlisting}

\begin{itemize}
	\item \lstinline{bool memento_info} - Flag for whether or not the root page is an archived page.
	\item \lstinline{string memento_datetime} - Datetime of the root page if it is a memento, otherwise set to "None".
	\item \lstinline{vector<string> memento_dates} - List of datetimes for archived elements on the page.
\end{itemize}

\begin{figure*}[ht]
\centering
\includegraphics[width=0.8\textwidth, frame]{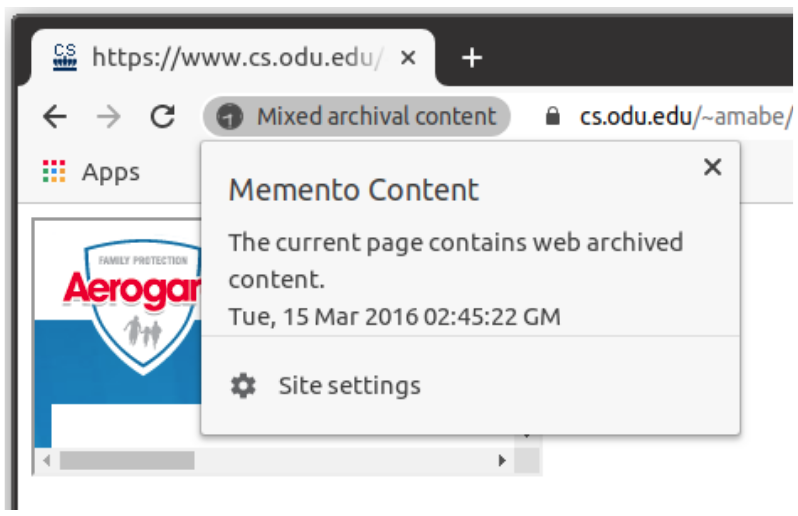}
\caption{When a single iframe memento is displayed on a live webpage the user is alerted that they are viewing a mix of live and archival content and the datetime of the archived resource is listed.}
\label{fig:one-iframe-ui}
\end{figure*}

\begin{figure*}[ht]
\centering
\includegraphics[width=0.8\textwidth, frame]{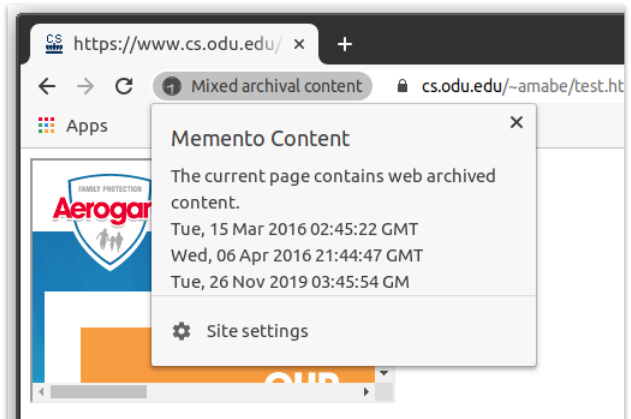}
\caption{When multiple iframe mementos are displayed on a live webpage the user is alerted that they are viewing a mix of live and archival content and the datetimes of all archived resources are listed.}
\label{fig:multi-iframe-ui}
\end{figure*}

With the list of datetimes on the page, \lstinline{memento_dates}, set within the \lstinline{NavigationEntry}, this list can be passed to the \lstinline{VisibleSecurityState} and later used to construct the proper UI elements. Figure \ref{fig:one-iframe-ui} shows the UI for when a single iframe memento is displayed on a live web page and Figure \ref{fig:multi-iframe-ui} shows when multiple iframe mementos are displayed on a live webpage.

\subsection{Live Web within Mementos}

\begin{figure*}[ht]
\centering
\includegraphics[width=0.52\textwidth, frame]{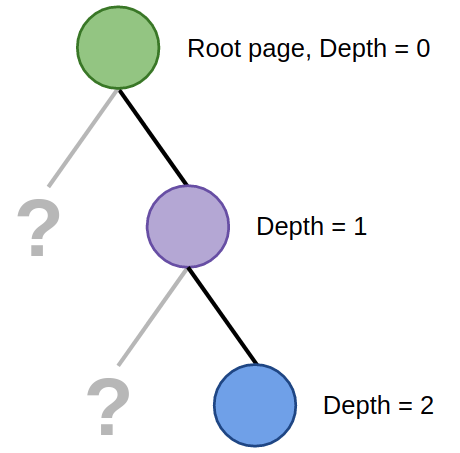}
\caption{Root page with an embedded frame that contains another embedded frame.}
\label{fig:height-3-tree}
\end{figure*}

When a page is loaded, all elements on the root page are loaded, including embedded elements within the embedded elements. Consider a webpage displaying an iframe, and that iframe displaying another iframe. The first iframe is a child node of the root page, and the second iframe is a child node of the first iframe. When the tree structure of the page is considered, this means the root page is the root node, the first iframe has a depth of 1, and the second iframe has a depth of 2. As shown in Figure \ref{fig:height-3-tree}, the nodes in the tree could also have any number of other child nodes aside from the ones described, meaning each frame could have any number of its own embedded frames.

The browser considers all these child nodes when constructing the \lstinline{VisibleSecurityState} for the whole page. As was shown in Figure \ref{fig:chromium-page-structure}, an insecure node at any depth of the tree could raise a security alert. The implementation described in Section \ref{section:iframe-elements-section} allows for datetimes returned by embedded elements at any depth to be collected. However, when a memento element is encountered its child nodes should be considered differently than regular child nodes. This is because children of the memento should return the same datetime as the memento itself. If the child nodes return no datetime, the memento is not correctly displaying content from the datetime it has returned and the live web is leaking into the archived element. This would create a webpage that never truly existed at the datetime returned by the resource, meaning that the user should be alerted that although it appears they are looking at an archived page from a certain date and time, the page never truly existed. Since the live web is being displayed within the archived page, this means a ``zombie" webpage is being displayed and the UI elements of the browser should alert the user that there is a mix of live and archival content on the page. This occurrence was described in Section \ref{section:embedded-elements-within-mementos} as one of the possible cases of embedded elements displaying within a memento. To account for this, the implementation of the \lstinline{Navigator} class was extended. If a currently processing child node of a memento does not have a datetime associated with its \lstinline{URLResponseHead} object, then the information for the root page needs to be updated so that it is known by the front-end that the page is displaying a mix of live and archival content (Listing \ref{lst:mixed-memento-flag}). Now the \lstinline{NavigationEntry} holds another variable for the memento information on the page:

\begin{itemize}
	\item \lstinline{bool memento_info} - Flag for whether or not the root page is an archived page.
	\item \lstinline{string memento_datetime} - Datetime of the root page if it is a memento, otherwise set to "None".
	\item \lstinline{vector<string> memento_dates} - List of datetimes for archived elements on the page.
	\item \lstinline{bool mixed_memento_live_web} - Flag for whether or not the root page is a memento that is displaying the current web or contains a memento that is displaying the current web.
\end{itemize}

\begin{lstlisting}[language=C++, caption=When a child frame of a memento does not return a datetime set the \lstinline{mixed_memento_live_web} flag for the root., label=lst:mixed-memento-flag, ]
  // Set the mixed_memento_live_web flag
  root->SetIsMixedMementoLiveWeb(true);
\end{lstlisting}

\begin{figure*}[ht]
\centering
\includegraphics[width=0.6\textwidth, frame]{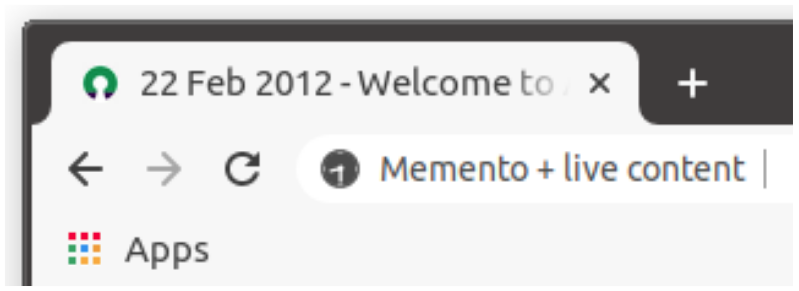}
\caption{Memento icon and message for when an archived element is displaying the live web.}
\label{fig:mixed-memento-UI}
\end{figure*}

With the \lstinline{mixed_memento_live_web} information for the root page appropriately updated, this can then be passed on to the \lstinline{NavigationEntry} so that the \lstinline{VisibleSecurityState} can have the information. Passing the information on to the \lstinline{VisibleSecurityState} allows the front-end to update as described in Figure \ref{fig:rootpage-loading-frontend}. The user will be presented with the memento icon as well as a message stating ``live + archival content" (Figure \ref{fig:mixed-memento-UI}).

\subsection{Memento Detection Evaluation}

Root page memento detection was tested by loading archived pages from a variety of web archives. Web archives may display mementos differently, however the browser should still react the same by updating the UI to display the memento icon and datetime. The root page memento detection was tested with the following web archives:

\begin{itemize}
    \item Archive-It \cite{archive-it}
    \item Archive.today \cite{archivetoday}
    \item Australian Web Archive (Trove) \cite{trove}
    \item BAnQ \cite{banq}
    \item Bibliotheca Alexandrina Web Archive \cite{bibalex}
    \item Icelandic Web Archive \cite{icelandic}
    \item Internet Archive \cite{internet-archive}
    \item Library and Archives Canada \cite{canada}
    \item Library of Congress \cite{congress}
    \item National Records of Scotland \cite{scotland}
    \item Perma Archive (Perma.cc) \cite{permacc}
    \item Portugeuse Web Archive \cite{portugeuse}
    \item Stanford Web Archive \cite{stanford}
    \item UK National Archives Web Archive \cite{uk-national-webarchives}
    \item UK Parliament Web Archive \cite{uk-parliament}
    \item UK Web Archive \cite{uk-webarchive}
\end{itemize}

\begin{figure*}[ht]
\centering
\includegraphics[width=0.8\textwidth]{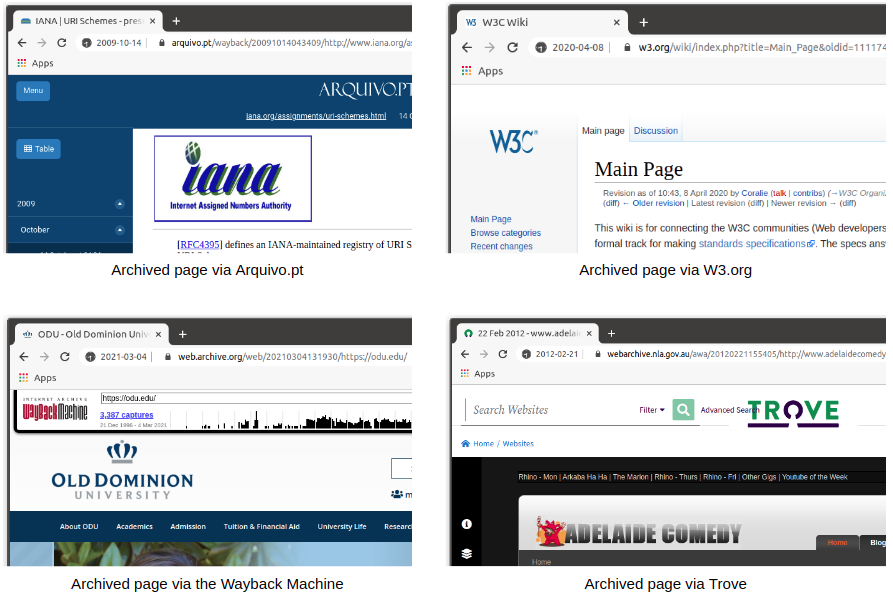}
\caption{Testing memento detection by loading archived pages from a variety of web archives.}
\label{fig:testing-webarchives}
\end{figure*}

Each of these web archives was navigated with the browser so that memento detection could be tested. While navigating the archive, the memento icon would not display until an actual archived page was selected, meaning the memento detection was working as expected by not displaying the icon while selecting a memento from the archive and then bringing up the icon when the memento was actually displayed in the browser tab. The testing of these web archives is shown in Figure \ref{fig:testing-webarchives}.

\subsection{Bookmark as Archive Implementation}
\label{section:bookmark-implementation}

\begin{figure*}[ht]
\centering
\includegraphics[width=0.6\textwidth, frame]{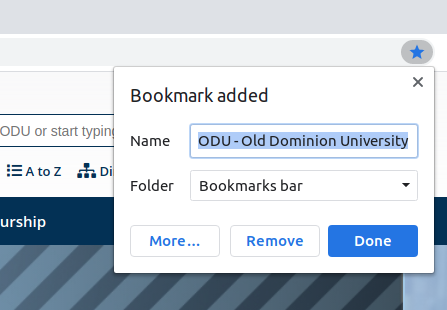}
\caption{The user can quickly hit the star icon to bookmark a URL. This creates a new bookmark node for that URL and brings up a popup where they can remove the bookmark or change its location.}
\label{fig:bookmark-popup}
\end{figure*}

The bookmark as archive feature was implemented in a way that fits with the current Chromium bookmarking. Normally when the user wants to bookmark a page, they would click the star icon in the upper right. This action immediately generates a \lstinline{BookmarkNode} for that URL and brings up a menu where the user can edit the bookmark options such as the name and location, or click ``Remove" to get rid of the bookmark immediately after they added it (Figure \ref{fig:bookmark-popup}). To add the archive dropdown, the implementation of the original bookmarking dropdown for selecting the location was copied to get started. The class that implements the location dropdown is \lstinline{RecentlyUsedFoldersComboModel}, and the title comes from the fact that the dropdown orders the bookmark locations based off of when they were most recently used. The version of this class for the archive dropdown is \lstinline{WebArchiveComboModel}. Changes in the class implementation between the two dropdowns occurs in the constructor which prepares the possible options that could be selected in the dropdown. The dropdowns may only display permanent bookmark nodes such as the ``Bookmarks bar" or the ``Other bookmarks" options. With the copy of the \lstinline{RecentlyUsedFoldersComboModel} class created to create the Web archive dropdown, this new class could be called alongside the old class from the \lstinline{BookmarkBubbleView} class so the new dropdown could be added with the original (Listing \ref{lst:new-dropdown}). 

This creates a second dropdown that is identical to the first. To put different options into the archive dropdown, the new options first need to be added to the implementation as permanent nodes. The options for this new dropdown are public web archives that can easily accept a submission: The Internet Archive, Archive.Today \cite{archivetoday}, and Megalodon.jp \cite{megalodon}. To add these web archives as permanent nodes, they were first added to the \lstinline{BookmarkModel} class header and the \lstinline{BookmarkNode} class header. Additionally, the option of ``No archive" was added so this would also appear in the dropdown (Listings \ref{lst:add-permanent-nodes} and \ref{lst:add-new-nodes}).

\begin{lstlisting}[float, language=C++, caption=Adding a second combobox to the add bookmark popup., label=lst:new-dropdown, ]
  // Add the original dropdown
  auto parent_folder_model = std::make_unique<RecentlyUsedFoldersComboModel>(
      model, model->GetMostRecentlyAddedUserNodeForURL(url_));

  // Add the archive dropdown
  auto parent_folder_model2 = std::make_unique<WebArchiveComboModel>(
      archive_model, archive_model->GetMostRecentlyAddedUserNodeForURL(url_));
\end{lstlisting}

\begin{lstlisting}[float, language=C++, caption=Create variables for the new permanent nodes within the bookmark model., label=lst:add-permanent-nodes, ]
  BookmarkPermanentNode* bookmark_bar_node_ = 
                                            nullptr;
  BookmarkPermanentNode* no_archive_node_ = 
                                            nullptr; // New permanent node
  BookmarkPermanentNode* archive_today_node_ = 
                                            nullptr; // New permanent node
  BookmarkPermanentNode* internet_archive_node_ = 
                                            nullptr; // New permanent node
  BookmarkPermanentNode* megalodon_node_ = nullptr; // New permanent node
  BookmarkPermanentNode* other_node_ = nullptr;
  BookmarkPermanentNode* mobile_node_ = nullptr;
\end{lstlisting}

\begin{lstlisting}[float, language=C++, caption=Add new permanent node types for the bookmark as archive dropdown., label=lst:add-new-nodes, ]
  enum Type {
    URL,
    FOLDER,
    BOOKMARK_BAR,
    NO_ARCHIVE, // New permanent node type
    ARCHIVE_TODAY, // New permanent node type
    INTERNET_ARCHIVE, // New permanent node type
    MEGALODON, // New permanent node type
    OTHER_NODE,
    MOBILE
  };
\end{lstlisting}

In addition to adding the new possible permanent types, they also needed their own unique GUIDs. Variables for this were added to the \lstinline{BookmarkNode} class header so that they could be given values within the class implementation (Listing \ref{lst:add-new-guids}).

\begin{lstlisting}[float, language=C++, caption=Add the variables for the archive node GUIDs and then values will be given within the \lstinline{BookmarkNode} class implementation., label=lst:add-new-guids, ]
  static const char kRootNodeGuid[];
  static const char kBookmarkBarNodeGuid[];
  static const char kNoArchiveNodeGuid[]; // New GUID
  static const char kArchiveTodayNodeGuid[]; // New GUID
  static const char kInternetArchiveNodeGuid[]; // New GUID
  static const char kMegalodonNodeGuid[]; // New GUID
  static const char kOtherBookmarksNodeGuid[];
  static const char kMobileBookmarksNodeGuid[];
  static const char kManagedNodeGuid[];
\end{lstlisting}

In addition to adding the new permanent nodes to the \lstinline{BookmarkModel} and the \lstinline{BookmarkNode} classes, references to the nodes also need to be added in various places within the \lstinline{BookmarkCodec} class. The \lstinline{BookmarkCodec} class handles encoding and decoding the bookmark nodes from the Chromium config file that contains the JSON information of the user's bookmarks. Anywhere the \lstinline{bookmarks_bar} permanent node was accessed or altered in the \lstinline{BookmarkCodec} class, the new permanent nodes were added and accessed or altered in the same fashion. After adding the permanent nodes to the implementation, they could be pushed into the list of nodes displayed in the archive dropdown. This was done within the constructor for the archive dropdown in the \lstinline{WebArchiveComboModel} class (Listing \ref{lst:add-archive-options}).

\begin{lstlisting}[float, language=C++, caption=Push the new permanent archive nodes onto the stack of options that will be displayed in the archive dropdown., label=lst:add-archive-options, ]
  items_.push_back(Item(model->no_archive_node(), Item::TYPE_NODE));
  items_.push_back(Item(model->archive_today_node(), Item::TYPE_NODE));
  items_.push_back(Item(model->megalodon_node(), Item::TYPE_NODE));
  items_.push_back(Item(model->internet_archive_node(), Item::TYPE_NODE));
\end{lstlisting}

\begin{figure*}[ht]
\centering
\includegraphics[width=0.6\textwidth, frame]{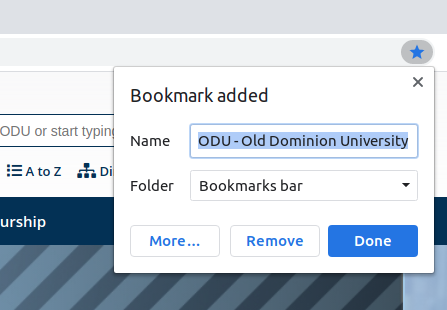}
\caption{The archive dropdown appears in the popup that displays when the user clicks the star icon to bookmark a page.}
\label{fig:archive-dropdown}
\end{figure*}

At this point in the implementation of the bookmark as archive feature, the archive dropdown was successfully displayed within the bookmark popup and the user could view the different web archive options, which are ``None", ``Archive.today", ``Internet Archive", and ``Megalodon" (Figure \ref{fig:archive-dropdown}). Clicking ``Done" to save changes in the popup does not do anything with the selected option in the archive dropdown. To implement this, the \lstinline{MaybeChangeParent} function within the \lstinline{WebArchiveComboModel} must be modified. This function is called when the ``Done" button is clicked as it is intended to ``maybe change the parent" of the bookmark depending on if the user changed its folder location. The bookmark folder dropdown class will keep this implementation of possibly changing the parent folder of the bookmark, but the archive dropdown class will need to submit the bookmarked URL to the selected public web archive. To submit to a public web archive, the browser can use ArchiveNow \cite{jcdl18:archivenow}, a tool to push web resources into web archives \cite{archivenow}. ArchiveNow is created with Python3, so the Chromium C++ implementation needs to call Python code in order to use ArchiveNow. This can be done using \lstinline{system} to make a system command line call to the Python script. However, this \lstinline{system} call will need to be different depending on the operating system. The command can be changed according to the operating system using the same technique that is shown in Listing \ref{lst:os_spec_header_files}. To construct the command, we need:

\begin{itemize}
	\item The location of the current working directory
	\item The title of the archive to submit to
	\item The URL of the webpage the user wants to submit
\end{itemize}

The location of the current working directory can be found with the function shown in Listing \ref{lst:cur-directory}.

\begin{lstlisting}[float, language=C++, caption=Function to get the path of the current working directory., label=lst:cur-directory, ]
  std::string get_current_dir() {
	char buff[FILENAME_MAX]; //create string buffer to hold path
	(void)GetCurrentDir( buff, FILENAME_MAX );
	std::string current_working_dir(buff);
    return current_working_dir;
  }
\end{lstlisting}

The title of the selected archive is the title of the bookmark node the user selected in the dropdown, so this can be easily accessed in the following way:

\begin{lstlisting}[language=C++, caption=Access the title of the bookmark node the user selected from the archive dropdown., label=lst:selected-archive, ]
  items_[selected_index].node->GetTitle()
\end{lstlisting}

The URL of the webpage the user wants to archive can be accessed through the bookmark node that was created when the user clicked the star icon:

\begin{lstlisting}[language=C++, caption=Access the URL of the page the user just bookmarked and wants to archive., label=lst:archiving-url, ]
  node->url().spec()
\end{lstlisting}

On macOS and Linux, the final Python command to call the script that utilizes ArchiveNow will look as follows: \lstinline{python3} \lstinline{path/to/script.py} \lstinline{'Internet Archive'} \lstinline{'https://example.com'} \lstinline{ &}. The ``\&" runs the Python script in the background. The command will look similar on Windows but the method of telling the script to run in the background is different. With the script to utilize ArchiveNow called, the ``Done" button in the bookmark popup is successfully submitting the bookmarked URL to be archived. However, this does not mean the user can access the archived page from the browser at this point. The star icon creates a standard bookmark node for the URL when clicked, and if the user has not selected ``None" in the archive dropdown then when they click the done button, the URL will be submitted to their selected web archive. This implementation so far has not created a node for the archived version of the webpage. The node for the archived webpage is also unable to have the proper URL to the page in the Web archive until it is done archiving, and archiving a page can take anywhere from a few seconds to several minutes. To mitigate this, the original plan was to add an archived bookmark node that links to the original page and the title of the node says that it is still archiving, then update the URL of that node in the background with the URL to the archived page once it is done archiving. Adding the placeholder archived bookmark node was simple and was done as shown in Listing \ref{lst:add-archived-node} within the \lstinline{MaybeChangeParent} function of the \lstinline{WebArchiveComboModel} class.

\begin{lstlisting}[language=C++, caption=Add a placeholder bookmark node for the archived page with the intent to update the URL with the URL to the page in the Web archive once it has completed archiving., label=lst:add-archived-node, ]
  bookmark_model_->AddURL(new_parent, new_parent->children().size(), base::UTF8ToUTF16(std::string("Archiving " + node->url().spec())), node->url().spec());
\end{lstlisting}

In Listing \ref{lst:add-archived-node}, \lstinline{node}\lstinline{->}\lstinline{url()}\lstinline{.}\lstinline{spec()} is the string URL of the bookmarked page so, for example, the title of the archived page bookmark node becomes something like ``Archiving https://example.com". And the URL that the archive bookmark node leads to is initially set to be the URL to the original page. When the page has finished archiving, the following parts of the archive bookmark node need to be updated:

\begin{itemize}
	\item The title of the node needs to be updated from ``Archiving <URL>" to something such as ``Archive.today example.com 2020-03-04".
	\item The URL that the node leads to needs to be updated to the URL to the archived version of the webpage.
\end{itemize}

\begin{figure*}[ht]
\centering
\includegraphics[width=0.9\textwidth, frame]{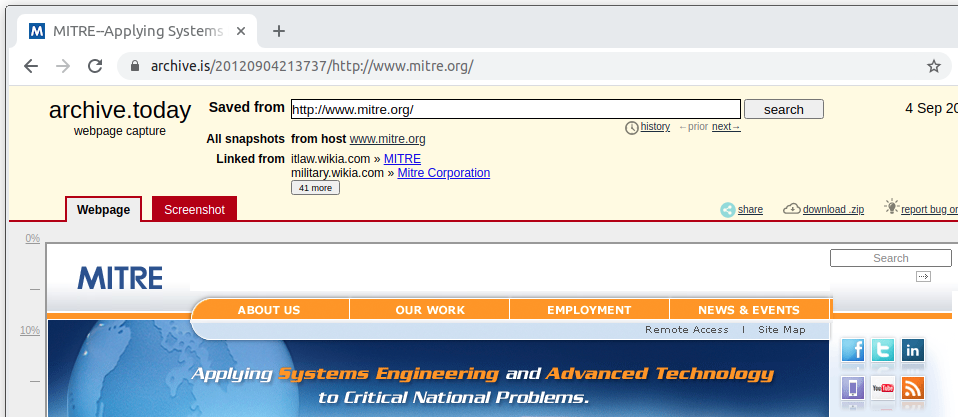}
\caption{If you try to load a memento of mitre.org from Archive.today from the datetime 20100412125057, it will redirect to a memento from 2012 since that is the closest date Archive.today has to the datetime given.}
\label{fig:mitre-2012}
\end{figure*}

However, updating the bookmark node proved to be very difficult. Any interaction with the bookmark nodes runs on the UI thread, meaning that the front-end updates immediately whenever something happens with the bookmarks. This is why when the user bookmarks a page and places it on the bookmarks bar, the browser immediately displays the new bookmark node on the bookmarks bar. Since all bookmark actions run on the UI thread and the archiving runs on the background thread since it can take several minutes, when the archiving is complete the bookmark node cannot be updated without crashing the browser with an invalid sequence error. Since this issue could not be quickly resolved, an alternative was implemented. Instead of creating a placeholder bookmark node that links to the actual webpage since it is expected to be updated, a bookmark node that links to that webpage in the archive at the time the ``Done" button was clicked is created. The Internet Archive's Wayback Machine and Archive.today both support URLs with 14 digit date strings. For example, the URL \href{https://web.archive.org/web/20100412125057/http://www.mitre.org/}{web.archive.org/web/20100412125057/http://www.mitre.org/} links to an archived webpage in the Wayback Machine. If you change the datestring to one for which the Wayback Machine does not have a memento, it will redirect to the closest datetime that they do have. Meaning if you put the datestring for the current date and time, it will take you to the latest memento for that webpage. The same is true for Archive.today, if you try to load the same URL for a memento of mitre.org but change the hostname to archive.is, it will take you to Archive.today's closest memento to that date: \href{https://archive.is/20100412125057/http://www.mitre.org/}{https://archive.is/20100412125057/http://www.mitre.org/}. That URL will actually load a memento of mitre.org in Archive.today from 2012, so the closest memento Archive.today has of mitre.org to 2010 is from 2012 (Figure \ref{fig:mitre-2012}).

Since the Internet Archive's Wayback Machine and Archive.today support the 14 digit datetime format and will redirect to the memento closest to that date, an archive bookmark node can be created with a working URL without waiting for the page to be archived. The URL will be constructed of the hostname for the chosen web archive, the 14 digit datetime string for the time the user submitted bookmark edits, and the URL of the page the user wanted to archive. This constructed URL will take the user to the memento in the archive closest to that date, which initially will be the latest memento for that URL, but once the page is finished archiving again from the script initiated by clicking the ``Done" button, the datetime in the URL will be closest to that memento so that is where it will redirect to. For example, if on March 4th, 2021 at 3:00pm the user clicks the ``Done" button to submit the website example.com to be archived by the Wayback Machine, the 14 digit datetime string will be 20210304030000. The URL for the archive bookmark node will be https://web.archive.org/web/20210304030000/https://example.com. If the webpage finishes archiving two minutes later at 3:02pm, then the true 14 digit datetime string for the archived page will be 20210304030200. Since the datetimes only have a two minute difference, the datetime for when the user clicked the button at exactly 3:00pm will redirect to the memento that was archived at 3:02pm. This workaround cannot be done for Megalodon.jp since it does not redirect to the nearest datetime. All archived page URLs are additionally printed to the file \lstinline{archive_urls.txt} so that they can be accessed once the page finishes archiving even though they cannot currently be added to the browser bookmarks.

\begin{figure*}[ht]
\centering
\includegraphics[width=0.8\textwidth, frame]{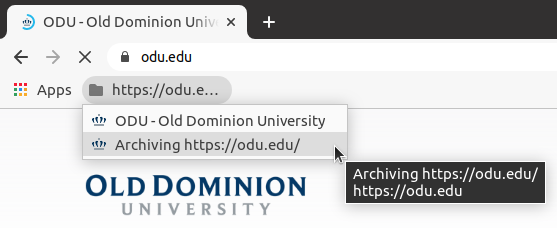}
\caption{When the user archives a page they have bookmarked, a folder gets created to hold all the bookmark nodes that lead to that webpage, either live or archived.}
\label{fig:archived-bookmark-folder}
\end{figure*}

In addition to creating the original bookmark node and the archive bookmark node, the browser also needs to place each node into a location that makes sense. When the user initially clicks the star icon to bookmark a page, the browser will create the node for that page and place it into the last bookmark folder used. Typically, this is the bookmarks bar. The archive bookmark could be placed into this location alongside the original bookmark, but the user has the option to archive the page again from the edit bookmark popup, thus creating more archive bookmark nodes. The original bookmark and the archive bookmark nodes should, by default, be placed into a location together since they all lead to the same webpage. That way the user can easily choose if they want the live page or one of the archived versions they saved. However since the user can archive a page multiple times and create multiple archive bookmark nodes, all these nodes must be managed and kept together efficiently. To implement this, when the user chooses to archive a webpage when they bookmark it, when the archive bookmark node is created a folder will also be created. The original bookmark node will be placed into this folder alongside the archive bookmark node. The title of the folder will be the URL to the live webpage. This way, the user will see their single bookmark folder node on the bookmarks bar (or wherever they chose to save the original bookmark) and when they click it they will see a dropdown showing all the possible versions of that page they saved (Figure \ref{fig:archived-bookmark-folder}).

To accomplish this, the implementation within the \lstinline{MaybeChangeParent} function of the \lstinline{WebArchiveComboModel} class was expanded (Listing \ref{lst:add-bookmark-folder}).

\begin{lstlisting}[language=C++, caption=Add a bookmark folder node to store the nodes for a specific webpage., label=lst:add-bookmark-folder, ]
const BookmarkNode* new_parent = 
  		bookmark_model_->AddFolder(
  				node->parent(), 
          node->parent()->children().size(), 
          base::UTF8ToUTF16(node->url().spec()));

bookmark_model_->Move(node, new_parent, new_parent->children().size());
\end{lstlisting}

In the listing, a new folder is added and set as the \lstinline{new_parent} variable. Next, the original bookmark node is moved into this \lstinline{new_parent}, meaning that the newly created folder becomes the parent of the original bookmark node. The only issue with this implementation is that a new folder will be created each subsequent time that webpage is archived. To mitigate this, the code can be wrapped in a condition where it only creates that new folder if it has not already been created and set as the parent of the original bookmark node (Listing \ref{lst:conditionally-add-bookmark-folder}).

\begin{minipage}{\linewidth}
\begin{lstlisting}[language=C++, caption=Wrap the bookmark folder node in a condition so that it does not execute if the page is being archived again., label=lst:conditionally-add-bookmark-folder, ]
if (items_[selected_index].node->GetTitle() !=
    base::UTF8ToUTF16(std::string("None"))) {

	const BookmarkNode* new_parent = 
	  		bookmark_model_->AddFolder(
	  			   node->parent(), 
	         node->parent()->children().size(), 
	         base::UTF8ToUTF16(node->url().spec()));

	bookmark_model_->Move(node, new_parent, new_parent->children().size());

}
\end{lstlisting}
\end{minipage}

\subsection{Bookmark as Archive Evaluation}

Since the workaround for the bookmark as archive URLs does not link to the exact mementos and instead depends on being redirected to the nearest, there is a chance it will link to a memento that is not the one created from the user's submission. Since archiving takes some time, it is also possible that there is a reasonable offset that can be added to the 14 digit date string of when the user clicks the ``Done" button. For example, if a page typically takes 60 seconds to archive, then 60 seconds could be added to the 14 digit date string in order to increase the chance the URL will redirect to the exact correct memento.

To evaluate this, 17 timemaps were analyzed to see how often the mementos in the timemap are very close together. The 17 timemaps were of popular news websites \cite{nwala2020365} and were chosen since they are likely to have been archived frequently. Additionally, only the portions of the timemaps from 2021 were considered since earlier portions of the timemaps (such as from around 2000) are likely to not have many mementos.

For each timemap, a datetime for each second between January 1, 2021 and the current date was generated and the memento closest to that datetime was found. Then, an offset of 30 seconds was added to the datetime and the memento closest to that new datetime was found. Next, these two mementos were compared. It was found that 99.1\% of the time the memento closest to the random datetime and the memento closest to that random datetime plus 30 seconds were the same. A 60 second offset was tested and the mementos were the same 98.4\% of the time. With a 120 second offset, the accuracy was 96.9\%. In this evaluation, only seconds were considered since the memento datetimes have second granularity.

Since webpages always take some time to archive, if we assume this to be around 60 seconds, then adding a 60 second offset to the datetime used to construct the URL should lead to the correct memento about 98\% of the time.

\begin{figure*}[ht]
\centering
\includegraphics[width=0.7\textwidth, trim=0 0 375 0, clip, frame]{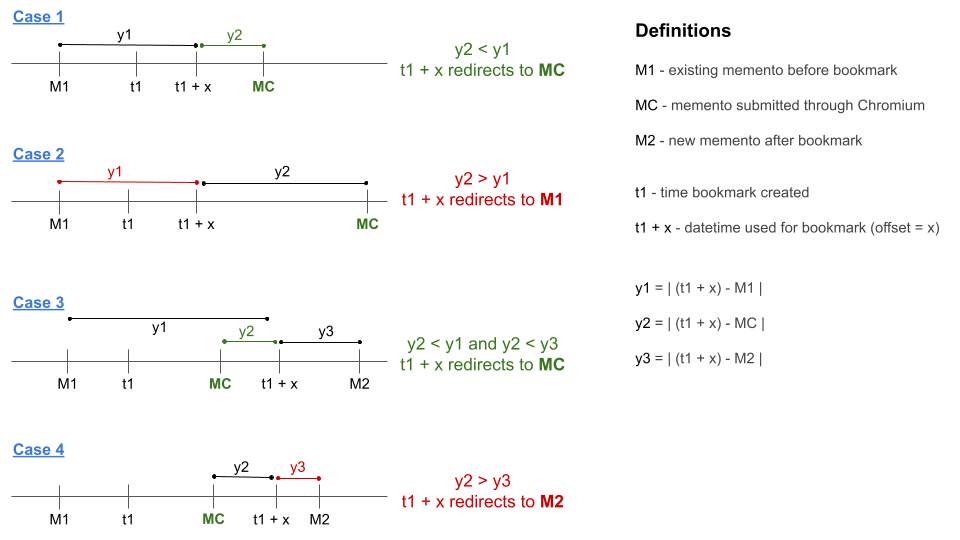}
\caption{Four cases of timelines for mementos and offsets for ``bookmark as archive".}
\label{fig:bookmark-eval}
\end{figure*}

The timeline for mementos and offsets is shown in Figure \ref{fig:bookmark-eval}. There are four possible cases, two that are successful (Case 1 and Case 3) and two that result in linking to an incorrect memento (Case 2 and Case 4). In the figure, \textsf{M1} stands for the latest memento before the bookmark is created, \textsf{t1} is the time the bookmark request is made, \textsf{t1+x} is the datetime used for the bookmark where \textsf{x} is the offset value we are testing, and \textsf{MC} is when the bookmark memento is actually created. The difference between Case 1 and Case 2 is how long it takes for the archive to create the new memento.  Just looking at Cases 1 and 2, we might think it best to set the offset \textsf{x} to be large.  But the cases (Cases 3 and 4) where another memento, \textsf{M2}, is created in the archive (by some other process) after the bookmarking process has started must also be considered. If the offset results in the bookmark node URL being too close to \textsf{M2} (Case 4), then the bookmark node will link to the incorrect memento. 
The archive will redirect to the closest existing datetime, so we want to set \textsf{t1+x} so that in most cases, it is closer to \textsf{MC} than any other mementos, without knowing exactly how long the archive will take to capture the page or what other mementos exist. The current feature implementation applies no offset, however from this evaluation a 30 second or 60 second offset would be reasonable to add.

\section{Usage}

This section first describes the browser from the user's perspective and what memento-aware features they can use. It also describes what partially implemented features can be seen from the UI. Also described is how to run the browser from the source code on GitHub \cite{repo} on both Linux and Windows, as well as instructions for switching between generating a debug build or a release build. The source code for the Memento-aware Browser on GitHub can only be built and run on Linux and Windows.

\subsection{System Walkthrough}
\label{section:system-walkthrough}

\begin{figure*}[ht]
\centering
\includegraphics[width=0.7\textwidth, frame]{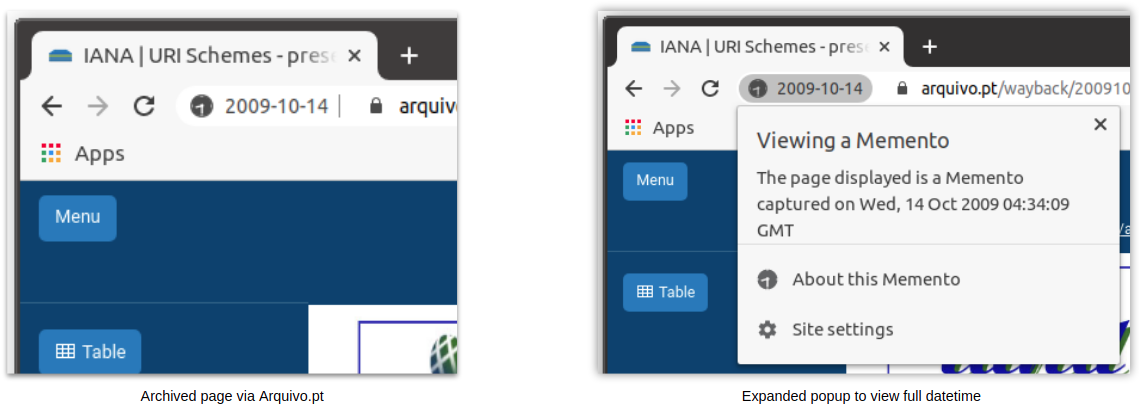}
\caption{If the root page is considered to be an archived page, the memento icon will display along with the datetime in YYYY-MM-DD format and the user may click the icon to bring up a popup with more information.}
\label{fig:final-memento-detection}
\end{figure*}

The main feature of the Memento-aware Browser is memento detection. If the user visits a webpage that is archived, they should expect to see the memento icon plus the datetime in YYYY-MM-DD format to appear to the left of where the HTTPS secure lock icon appears. Figure \ref{fig:final-memento-detection} shows an archived page with the memento icon visible and the expanded popup.

\begin{figure*}[ht]
    \centering
    \begin{tabular}{cc}
        \includegraphics[width=70mm]{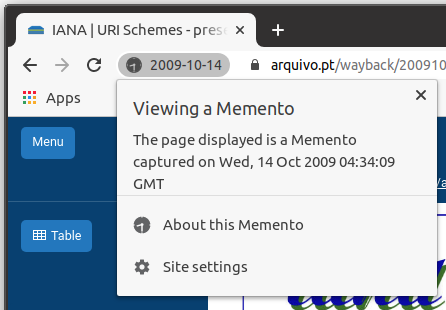} &   \includegraphics[width=90mm]{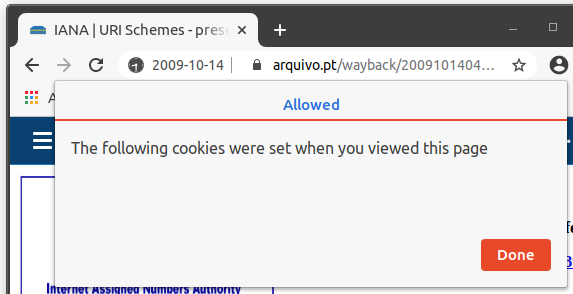} \\
        (a) memento info popup & (b) About this memento blank popup \\
    \end{tabular}
    \caption{The memento info popup appears when the user clicks the memento icon and the blank popup appears when they click ``About this memento"}
    \label{fig:memento-popups}
\end{figure*}

Additionally, just like the user can click the HTTPS secure lock icon to bring up the connection popup for more connection security information, they can also click on the memento icon for more information about the archived page. The current browser implementation brings up a popup very similar to the connection popup, except the text is about the webpage being archived. As shown in the left hand side of Figure \ref{fig:memento-popups}, the popup states that the user is currently viewing a memento, and that the page being displayed by the current browser tab is an archived page that was captured at a certain date and time. Additionally, there are two buttons on the popup. The first button says ``About this Memento" and the second is the standard ``Site settings" button that also appears on the connection popup. The ``About this Memento" button leads to a currently blank popup (right hand side of Figure \ref{fig:memento-popups}) that is further discussed in \ref{section:future-enhancements-section}. As for the ``Site settings" button, it is there by default with the popup and was not removed when the connection popup was copied to create the memento info popup since it could be used in the future for settings regarding archived content.

\begin{figure*}[ht]
\centering
\includegraphics[width=0.6\textwidth, frame]{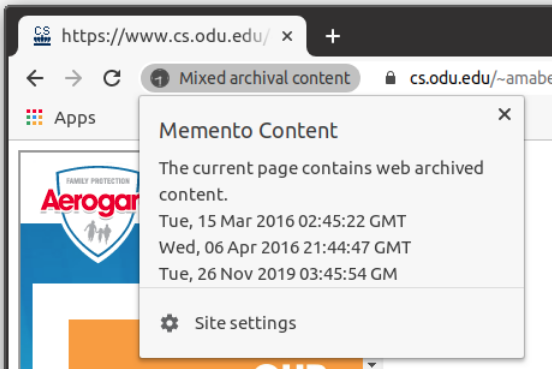}
\caption{The popup will display all three datetimes that exist on the page.}
\label{fig:three-mementos-UI}
\end{figure*}

The next part of the memento detection feature that the user can experience is the browser detecting archived content within a live webpage. For this, two example webpages are available where the first contains one memento (\url{https://www.cs.odu.edu/~amabe/oneiframe.html}) and the second contains three (\url{https://www.cs.odu.edu/~amabe/test.html}). The first example webpage is a live webpage with an iframe element that is displaying a memento. This scenario is supposed to simulate a live webpage displaying archived content as a part of the page, and not the whole page. For example, an article on the live web may want to show an embedded memento so that their readers can see the archived content they are referring to. The second example webpage is similar, except there are three embedded mementos instead of one. When the user visits the first example webpage they can expect the Memento-aware Browser to display the memento icon, but instead of displaying a datetime next to the icon, it will show the message ``Mixed archival content". When the user visits the second example webpage, they can expect the browser to display the memento icon and message just as it did for the first example page, but the popup will list three datetimes, one for each archived frame on the page. This is shown in Figure \ref{fig:three-mementos-UI}.

\begin{figure*}[ht]
\centering
\includegraphics[width=0.6\textwidth, frame]{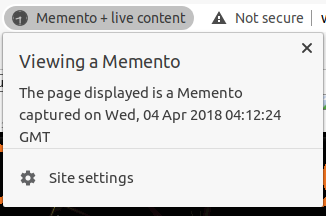}
\caption{The browser alerts the user that the archived page they are viewing is displaying content from the live web but also allows them to see the datetime the archived page is supposed to be from.}
\label{fig:memento-plus-live}
\end{figure*}

The final piece of the memento detection feature the user can expect is the detection of ``zombies" \cite{zombies}, or archived pages that are displaying content from the live web. With this feature, when an archived page is visited, the browser displays the memento icon as usual and detects if any of the frames being displayed by that archived page are from the live web. If live content is detected in the archived page, instead of showing the datetime by the memento icon, the browser will display the message ``Memento + live content". Additionally, the popup that appears when the user clicks the memento icon will display the same message as usual when an archived page is displayed (Figure \ref{fig:memento-plus-live}).

\begin{figure*}[ht]
\centering
\includegraphics[width=0.58\textwidth, frame]{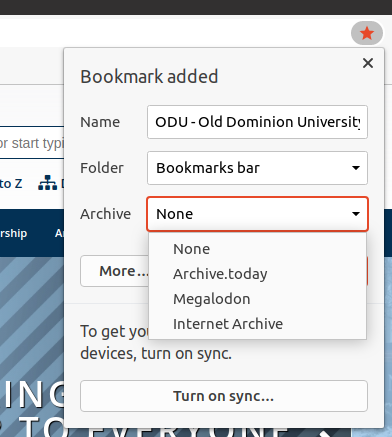}
\caption{The bookmark as archive option appears in the edit bookmark popup.}
\label{fig:archive-options}
\end{figure*}

In addition to the memento detection, the user may also use the bookmark as archive feature. When bookmarking a page through the star icon, the edit bookmark popup will appear which allows the user to change the name of the bookmark, change the location, or submit that page to a public web archive. To archive the page from this edit bookmark popup, the user may select a web archive from the archive dropdown and then either click the ``Done" button to submit or click out of the window. To exit the window without applying edits made to the bookmark, the user must click ``Cancel" to cancel all edits or ``Remove" to remove the bookmark entirely. Essentially, all edits made in the edit bookmark popup are immediately applied when the user clicks out of the window or clicks ``Done" which is standard in the Chromium implementation. With the archive dropdown being present on top of the original implementation, this means that the page is immediately submitted to the selected archive when the user clicks out of the popup or clicks the ``Done" button. Figure \ref{fig:archive-options} shows the edit bookmark popup with the newly added archive dropdown and its options. The available archives are the Internet Archive, Archive.today, and Megalodon.jp.

\begin{figure*}[ht]
\centering
\includegraphics[width=0.99\textwidth, frame]{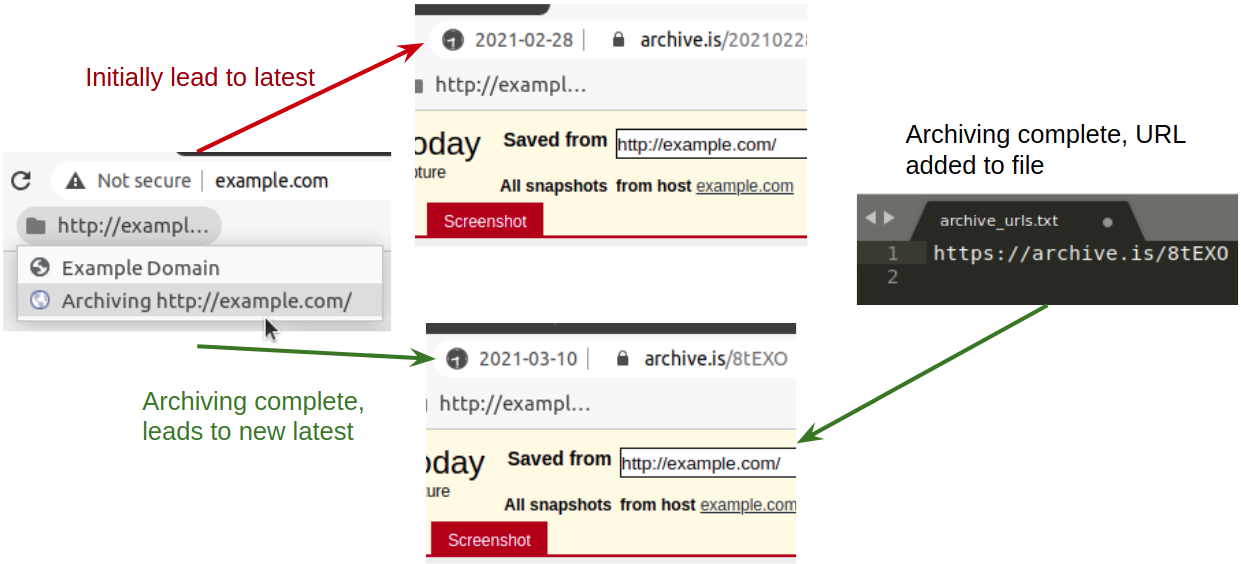}
\caption{The archive bookmark node will redirect to the memento in the archive that is closest to the datetime the page was submitted from the edit bookmark popup. Initially, this may be an older memento.}
\label{fig:bookmark-as-archive-ux}
\end{figure*}

When the edit bookmark popup changes are applied and the user has selected a web archive, they can expect the browser to create a bookmark node folder for that webpage and place the original bookmark for the page into that folder. The browser will also create an archive bookmark node and place it into the folder alongside the original bookmark node. If the user had selected either the Internet Archive or Archive.today, then the archive bookmark node that is created will link to an archived page in the chosen web archive from the datetime that the bookmark edits were applied. If the user has selected Megalodon.jp, then the archive bookmark node will link to the live webpage. In the background, the user can expect the browser to be archiving the webpage. When it is finished archiving, the URL will be added to the file \lstinline{archive_urls.txt}. This way the user can always access the URL to the archived page even though the browser cannot update the archive bookmark node. The reason as to why the URL is set differently depending on the archive and the archive bookmark node cannot be updated is explained in Section \ref{section:bookmark-implementation}. To walk through the example shown in Figure \ref{fig:bookmark-as-archive-ux}, say the user wanted to use the bookmark as archive feature to submit example.com to Archive.today on March 10, 2021 at 4:18pm. The user can expect the browser to immediately create a bookmark node folder for the webpage and place two bookmark nodes in the folder. The first bookmark node in the folder would be the standard bookmark that leads to the live webpage and the second bookmark node would be the one for the archived version of the page. The URL for the archive bookmark node would have been constructed using the method described in Section \ref{section:bookmark-implementation} and would initially redirect to Archive.today's latest memento of example.com, which at that time would be from February 28, 2021. Eventually, when the page is done archiving a few minutes later, the archive bookmark node URL will redirect to the memento of example.com that was archived a few minutes after the user submitted it with the bookmark as archive feature at 4:18pm. Additionally, the user can also expect the URL to the memento to appear in the \lstinline{archive_urls.txt} file. Rather than redirecting, this URL will lead directly to the memento of example.com that was archived a few minutes after the user submitted it with the bookmark as archive feature at 4:18pm.

\subsection{Running from Source}

The Memento-aware Browser can be run from the source code on GitHub on either Linux or Windows. The README file of the GitHub repository walks through running the code on either operating system. There are instructions for running on macOS, however these do not work due to missing dependencies and the issue could be fixed in the future. When running the source code, by default a debug version will be generated. A release version can be built by adding \lstinline{is_debug=false} to the file \lstinline{out/Default/args.gn}. In order to run from source, it is also necessary to download \lstinline{depot_tools} as described in Section \ref{section:working-with-chromium}. It is recommended to have about 100GB of free space before attempting to build the browser from source. Additionally, it is recommended to have at least 8GB of RAM to build and run the browser from source. 

\subsection{Release Builds}

The GitHub repository includes release builds for Linux and Windows \cite{repo}. In each of these releases, the memento detection feature is available without any additional dependencies. As for the bookmark as archive feature, ArchiveNow \cite{archivenow} and its dependencies will need to be installed as instructed on the release page \cite{releases}.

\section{Future Enhancements}
\label{section:future-enhancements-section}

This section outlines additional features and enhancements to the Memento-aware Browser. The features mentioned can be built off of the existing features, specifically the memento detection feature since it stores the memento datetimes for all resources but is currently only considering the datetimes returned by the main URL.

\subsection{``About this Memento" Popup}
\label{section:about-this-memento-popup}

\begin{figure*}[ht]
\centering
\includegraphics[width=0.99\textwidth, frame]{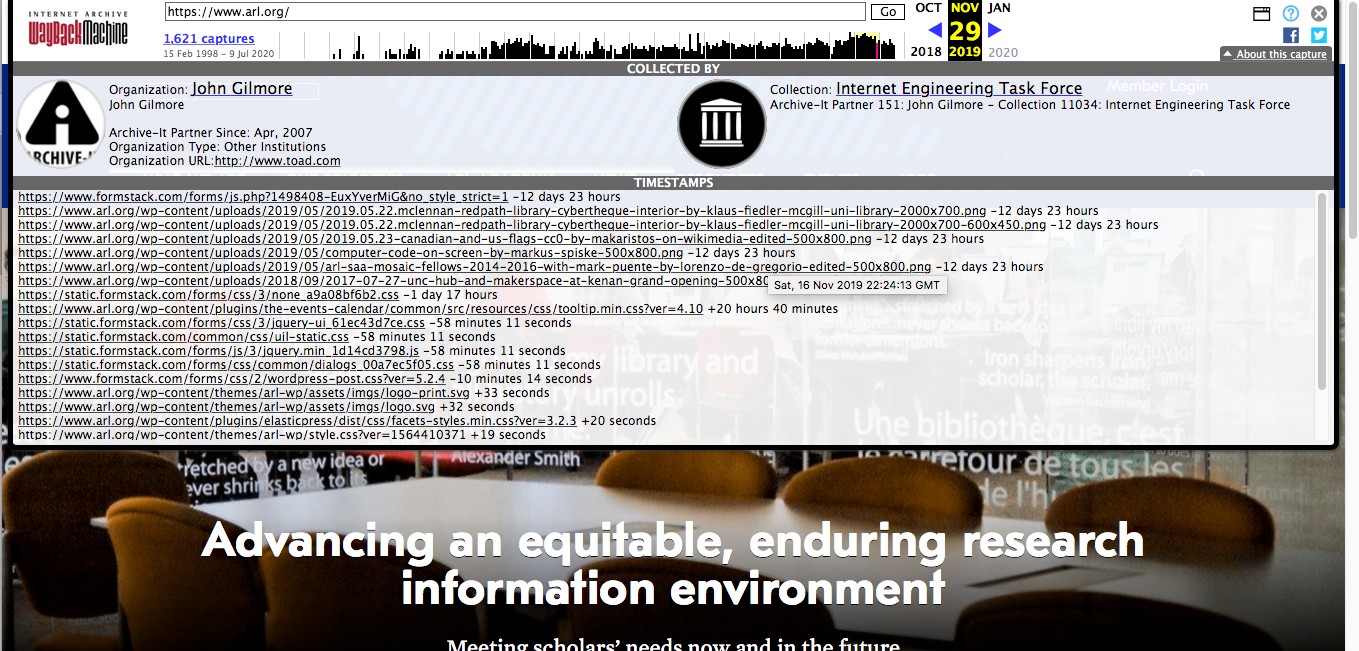}
\caption{The Internet Archive's Wayback Machine allows the user to see a list of archived resources that make up the archived page.}
\label{fig:about-this-capture}
\end{figure*}

The current memento detection provides information about whether or not the page displayed is an archived page as well as the datetime the page was archived. However, there is more information about the archived page that could be useful to the user. An example of this is a feature provided by the Internet Archive's Wayback Machine. The feature is a drop down labelled ``About this capture" that allows the user to see a list of all the archived resources and their respective datetimes that make up the entire webpage (Figure \ref{fig:about-this-capture}). Since the Memento-aware Browser stores the Memento-Datetime value for all resources as they are loaded as explained in Section \ref{section:detecting-memento-header}, implementing this feature would mostly consist of completing the front-end. Already in the browser is a button that would bring up the popup to display the information. The button is the ``About this memento" option that was detailed in Section \ref{section:system-walkthrough} and appears when the user clicks on the memento icon to get more information about the memento. Currently the button brings up an empty popup that was made from the collected cookies popup. If the feature were to be implemented, this popup would display a list of all the resources and their respective datetimes that make up the archived page. Additionally, it would be ideal if the user could click on a listing for a resource within the popup and that resource would highlight somehow within the browser tab. For example, if the user saw a listing for an image in the popup and they clicked that listing, the browser would highlight that image within the browser tab with a red box.

\subsection{List of Known Web Archives}

The current implementation of the memento detection just reacts to the presence of the Memento-Datetime HTTP response header and also the level the frame that returned the header exists within the page structure tree. It is possible for a live webpage to return the Memento-Datetime header. A possible way to account for this is for the browser to only acknowledge the header if the frame that is returning it is from a known web archive. This would involve the browser having a hard coded list of known and accepted web archives. Ideally, this list would be a part of the browser settings so that individual users could add or remove archives from the list stored in their local browser configuration. Initially the browser would come with its own list and then the user could customize it from there. A list of some possible web archives for the browser to recognize initially is the list of web archives the memento detection has been tested with. This list can be found within the docs section of the Memento-aware Browser GitHub repository \cite{repo-docs}.

\subsection{Temporal Coherence}

\begin{figure*}[ht]
\centering
\includegraphics[width=0.7\textwidth, frame]{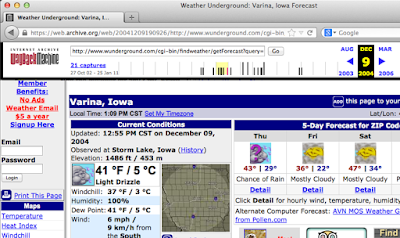}
\caption{Weather report archived on 2004-12-09 19:09:26 GMT \cite{ht15:as-presented}}
\label{fig:weather-report}
\end{figure*}

Another addition to the memento detection would be to consider the temporal coherence of the memento \cite{ht15:as-presented}. If you consider the HTML webpage and all the resources such as CSS and images required for the full and complete page, these pieces together can be defined as a composite memento \cite{ainsworth:composite:tr}. Just like the Chromium page structure defined in Section \ref{section:chromium-page-structure}, composite mementos can be thought of as a tree structure where the root resource (the main HTML webpage) has other resources connected to it. Consider an example of a weather report that was captured on December 9, 2004 at the time 19:09:26 GMT (Figure \ref{fig:weather-report}). The Memento-Datetime value returned by the page would be 2004-12-09 19:09:26 GMT, however if you look closer at the it would appear that the content does not make sense. The description of Current Conditions within the weather report states light drizzle, yet the rader shows no clouds. Looking at the datetimes for the page resources, it can be found that the radar image was captured 9 months later than the webpage itself, meaning that this weather report never actually existed on the live web. When something like this occurs, it would be ideal for the browser to alert the user that the page they are viewing is temporally incoherent and that although it appears they are looking at an archived webpage from a certain date and time, that webpage never actually existed as it is being presented.

\begin{figure*}[ht]
\centering
\includegraphics[width=0.74\textwidth, frame]{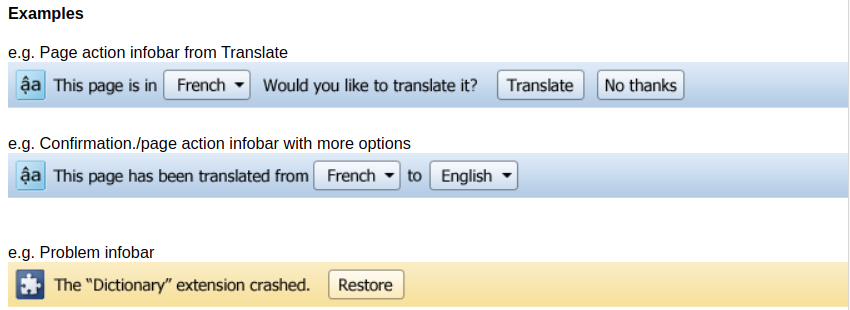}
\caption{Examples of infobar alerts that appear in the Chromium browser (\url{https://www.chromium.org/user-experience/infobars}).}
\label{fig:infobar-examples}
\end{figure*}

\begin{figure*}[ht]
\centering
\includegraphics[width=0.9\textwidth, frame]{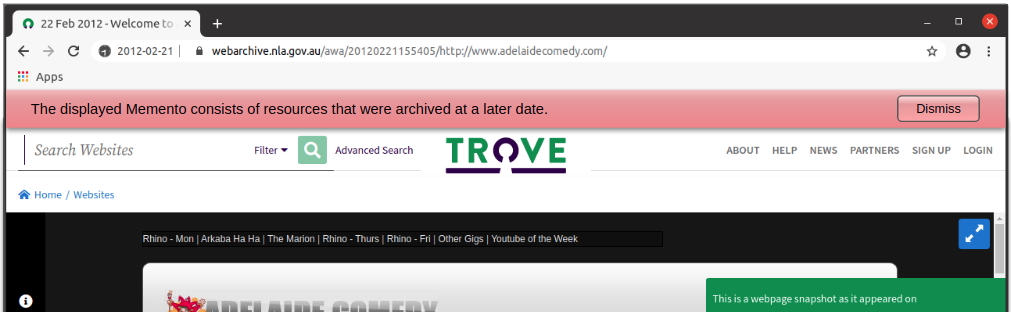}
\caption{Mock up of using an infobar to display an alert about temporal incoherence.}
\label{fig:infobar-UI}
\end{figure*}

As mentioned in Section \ref{section:about-this-memento-popup}, the Memento-aware Browser already stores the Memento-Datetime value for any resources that return it. Since the browser already has the necessary information about any composite mementos that are loaded, then the main part of the implementation for this feature would be classifying the different scenarios and presenting the appropriate UI elements. A suggested method to alert the user of a temporally incoherent memento is to display an infobar \cite{infobar}. Infobars appear in the Chromium browser for a variety of reasons. Some examples are shown in Figure \ref{fig:infobar-examples}. Many infobars displayed by the browser also have a ``Dismiss" button. So the user is alerted by the infobar appearing and once they have read and noted the information they can dismiss the popup. An infobar would be a great way to alert the user that the memento they are viewing is temporally incoherent since the memento icon will already be displaying datetime information. The user could also dismiss the warning since in some scenarios it may not matter to them that the archived page is temporally incoherent so they may want to view the page without any warnings. Figure \ref{fig:infobar-UI} shows an example of how this popup might look. The use of an infobar for temporal coherence information would also allow for different colored popups to be used depending on the situation, or even different options on the bar in addition to ``Dismiss". Another form of alert that the Chromium browser offers is a dialog box. However, a dialog box is not an optimal method of giving the user this information since dialog boxes should only be used if we want to block the user from doing something unless they take action, or the box appears in response to a user action. We do not want to force the user to acknowledge the temporal incoherence warning since it is not information that requires immediate action. Rather, it is additional information being offered to the user so they may keep this in mind when viewing the archived page.

\section{Conclusion}

A successful proof of concept Memento-aware Browser was created by extending Google's open source web browser Chromium \cite{chromium-download}. Throughout the process of adding the memento-aware features to the Chromium code base, it was found that these features work well within the native implementation. Detecting the Memento-Datetime HTTP response header proved to not be too difficult since the browser already needs to parse all response headers; it was mostly a matter of accounting for this header where the headers were already being parsed. Adding the basic memento detection described in Section \ref{section:basic-memento-detection} also worked with the native browser implementation very well. The memento detection features became more complex when different cases were accounted for, like the iframe memento case such as the one that can be see when viewing archived pages from Trove \cite{trove}. However once fully implemented, the user experience of the memento detection works well and makes sense when compared to the user experience when using the standard Chromium browser. The bookmark as archive feature implementation also fit into the native implementation well. However, there were challenges encountered while implementing this feature. It was found that since all bookmark actions run on the UI thread, bookmarks could not be updated on completion of the archiving that was running on the background thread. A workaround was implemented that utilized the fact that the Internet Archive and Archive.today redirect links that contain datetimes they do not have mementos for to the memento closest to that datetime.

The Memento-aware Browser source code is on GitHub \cite{repo} and can be built and run on Linux and Windows. Additionally, there are release builds for Linux and Windows attached to the GitHub repository. There are some existing issues with the browser, mainly in that the bookmark as archive feature is not considered complete. The implemented workaround for this feature only works for when the user selects the Internet Archive or Archive.today. Megalodon.jp \cite{megalodon} does not redirect links to the closest datetime so the workaround was not implemented for this archive. The memento detection feature works well but there likely are some corner cases that have not been accounted for. There is much room for improvement and additional features in the browser, however the current version serves as a great proof of concept for the user experience and the ability to work these features into the native browser implementation.

\clearpage

\bibliographystyle{ACM-Reference-Format}
\bibliography{refs}

\end{document}